\documentclass[a4paper,11pt]{article}
\pdfoutput=1 
\usepackage{jcappub}
\usepackage{graphicx}
\usepackage{epsfig,color}

\def\beq{\begin{equation}}
\def\eeq{\end{equation}}
\def\bey{\begin{eqnarray}}
\def\eey{\end{eqnarray}}

\def\lsim{\mathrel{\raise.3ex\hbox{$<$\kern-.75em\lower1ex\hbox{$\sim$}}}}
\def\gsim{\mathrel{\raise.3ex\hbox{$>$\kern-.75em\lower1ex\hbox{$\sim$}}}}

\title{Cluster Mergers and the Origin of the ARCADE-2 Excess}
\author[a]{Ke Fang}
\author[b]{, Tim Linden}
\affiliation[a]{Department of Astronomy, University of Maryland, College Park, MD, 20742, USA}
\affiliation[b]{Center for Cosmology and AstroParticle Physics, Ohio State University, Columbus, OH, 43210, USA}
\abstract{
Radio observations at multiple frequencies have detected a significant isotropic emission component between 22~MHz and 10~GHz, commonly termed the ARCADE-2 Excess. The origin of this radio emission is unknown, as the intensity, spectrum and isotropy of the signal are difficult to model with either traditional astrophysical mechanisms or novel physics such as dark matter annihilation.  We posit a new model capable of explaining the key components of the excess radio emission. Specifically, we show that the re-acceleration of non-thermal electrons via turbulence in merging galaxy clusters are capable of explaining the intensity, spectrum, and isotropy of the ARCADE-2 data. We examine the parameter spaces of cluster re-acceleration, magnetic field, and merger rate, finding that the radio excess can be reproduced assuming reasonable assumptions for each. Finally, we point out that future observations will   definitively confirm or rule-out the contribution of cluster mergers to the isotropic radio background.}


\begin{document}

\maketitle 

\section{Introduction}
\label{sec:introduction}

Observations of the extragalactic radio background have detected an isotropic excess in the frequency range of 22~MHz to 10~GHz~\citep{haslam_408Mhz_excess, roger_22MHz_excess, reich_1.4GHz_excess, guzman_44Mhz_excess, arcade_interpretation, arcade_measurement, 2014JCAP...04..008F}. The aggregate data, assembled by the Absolute Radiometer for Cosmology, Astrophysics, and Diffuse Emission-2 (ARCADE-2) Collaboration, show that the  emission has a spectrum that falls as T$^{-2.6}$~\citep{arcade_measurement}. The overall intensity of the emission (hereafter referred to as the ARCADE-2 excess), exceeds the intensity expected from the isotropic portion of the galactic diffuse emission by a factor of a few~\citep{arcade_interpretation, kogut_excess_not_galactic}.

Several models have been posited in order to explain the intensity of the ARCADE-2 signal, including contributions from radio galaxies, radio supernovae, radio-quiet quasars, and diffuse electrons in the inter-galactic medium~~\citep{singal_not_extragalactic_baryonic_signals}. However, the extrapolation of these source classes to low-luminosities has failed to reproduce the total intensity of the ARCADE-2 excess while maintaining consistency with constraints from infrared and X-Ray emission~\citep{gervasi_radio_luminosity_from_extrapolation_of_known_sources}. In addition to astrophysical sources, dark matter models have also been formulated in order to explain the ARCADE-2 excess. Dark matter models are well motivated for explanations of ARCADE-2 since they naturally produce large fluxes of relativistic e$^+$e$^-$ while producing negligible thermal emission. This allows dark matter induced emission to simultaneously fit the intensity of the ARCADE-2 excess while remaining consistent with infrared constraints, and several similar models have been formulated to date~\citep{fornengo_arcade_excess_is_dm, fornengo_arcade_excess_is_dm2, hooper_arcade_excess, FL14}.

However, a second peculiar feature of the ARCADE-2 excess pertains to the high isotropy of the extragalactic signal. It was first noted by Holder \citep{holder_anisotropy_of_arcade} that the high isotropy of the ARCADE-2 excess may be in tension with the small total-anisotropy observed in high angular resolution radio observations of the extragalactic sky. In fact, even making the conservative assumption that the ARCADE-2 emission dominates the total anisotropy observed in VLA and ACTA observations, Holder placed an upper limit on the ARCADE-2 anisotropy which falls below level expected for emission sources that trace large scale structure~\citep{holder_anisotropy_of_arcade}. Additionally, an analysis by Vernstrom et al.~\citep{Vernstrom:2014uda} calculated the luminosity distribution of detected radio sources at an frequency of 1.75~GHz, and found that a population of point sources would be unable to explain the total intensity of the ARCADE-2 emission unless they have individual emission intensities smaller than 1~$\mu$Jy. These observations strongly rule out most astrophysical, and dark matter source classes, requiring an extremely large number of very faint emission sources in order to produce the ARCADE-2 signal. 

One caveat common to both of the anisotropy constraints concerns the small angular field of view of radio interferometers. Due to the zero point subtraction necessary for this instrumentation, they are insensitive to emission which is isotropic on larger angular scales. Specifically, \citep{holder_anisotropy_of_arcade} constraints the total anisotropy level on angular scales ranging from $\ell$~=~6000 - 28000, while \citep{Vernstrom:2014uda} quotes a sensitivity to objects smaller than 2 arcminutes. Notably, this angular scale is smaller than that of the most massive galaxy clusters. In the context of dark matter models for the ARCADE-2 excess, it was noted by ~\citep{FL14} that the anisotropy from this emission may fall below the isotropy constraints of \citep{holder_anisotropy_of_arcade} and  \citep{Vernstrom:2014uda} in scenarios where synchrotron radiation from dark matter annihilation is dominated by galaxy clusters, with significant emission extending to several times the virial radius of these massive objects. 
While this model is intriguing, it requires large magnetic field extensions in cluster sized objects, compared to typical cluster models \citep{2002ARA&A..40..319C}. In addition, this model also requires large dark matter substructure boost factors, which  is consistent with the results of the Aquarius simulation  \citep{2008MNRAS.391.1685S, 2012MNRAS.419.1721G, 2012MNRAS.425.2169G}, but is disfavored by  \cite{2014MNRAS.442.2271S}.

An alternative model, first noted in~\citep{FL14}, concerns the re-acceleration of non-thermal electrons through hydrodynamic shocks produced in cluster mergers. In the context of the dark matter models originally considered by~\citep{FL14}, these shocks were employed to re-accelerate electrons originally produced via dark matter annihilation, significantly amplifying the total synchrotron flux produced via a dark matter annihilation event.  This mechanism is capable of smearing out the resulting synchrotron signal via re-acceleration, decreasing the total anisotropy of the synchrotron emission.
In this paper, we show that cluster mergers can independently produce the intensity, spectrum and morphology of the ARCADE-2 excess without the need for seed-electrons from dark matter annihilations. In this case, we note that any population of non-thermal seed electrons can be efficiently re-accelerated to GeV energies. Moreover, the cluster mergers also produce significant magnetic fields within the cluster, allowing the re-born electrons to produce significant synchrotron radiation in the ARCADE-2 band. Finally, since strong mergers are most prevalent in the most massive clusters, the anisotropy from this emission mechanism occurs primarily on scales larger than the anisotropy studies of \citep{holder_anisotropy_of_arcade, Vernstrom:2014uda}.
 Indeed turbulent reacceleration   is one of  the most popular models  employed to explain the observed radio emission at cluster scales~\citep{2014IJMPD..2330007B}.

The outline of the paper is as follows. In Section~\ref{sec:model} we discuss the physical model that produces synchrotron emission in cluster mergers, breaking down the process into (i) production of hydrodynamic turbulence and Alfv\'en waves, (ii) the Alfv\'en re-acceleration of e$^-$ and e$^+$ in the intra-cluster medium, (iii) profiles of the density and the magnetic fields of   clusters and the (iv) cluster merger rate. In Section~\ref{sec:results} we compare our resulting model for synchrotron radiation against the observable parameters of the ARCADE-2 excess. Finally in Section~\ref{sec:conclusions}, we discuss several caveats to our analysis and conclude.

\section{Models}\label{sec:model}

The production of the synchrotron emission in cluster mergers proceeds through several steps. Cluster mergers first induce fluid turbulence which then produces Alfv\'en waves. These Alfv\'en waves accelerate existing non-thermal electrons to GeV energies. The fluid turbulence also amplifies existing cluster magnetic fields, causing the GeV electrons to efficiently lose energy to synchrotron radiation in the GHz band.  We will discuss the assumptions and outputs of our model in each step of the synchrotron production process.

\subsection{Turbulence and Alfv\'enic waves}\label{subsec:turbulence}
In order to calculate the production of Alfv\'en Waves via fluid turbulence in cluster shocks, we closely follow the resonant acceleration model studied in \cite{1984ApJ...277..820E} and \cite{Fujita03}. Specifically, we assume that the energy spectrum of the turbulence induced by a cluster merger can be described by a power law:
\begin{equation}\label{eqn:W_f}
W_f(\kappa)=W_f^0\,\left(\frac{\kappa}{\kappa_0}\right)^{-m}
\end{equation}
where $W_f$ is the energy density of the turbulent fluid at a wavenumber $\kappa$, where $\kappa=2\pi/l$ is the wavenumber for a length scale $l$. The value $\kappa_0=2\pi/l_0$ corresponds to the largest eddy size of the system, $l_0$.  The energy density of the turbulence is given by $E_t\sim \rho v_t^2$, where $\rho$ is the mass density of the fluid and $v_t$ is the turbulent velocity at the largest scale $l_0$. Noting that the total energy density is the sum of the scales $\kappa$, we have $E_t = \int W_f(\kappa)d\kappa$ and thus:
\begin{equation}\label{eqn:Wf0}
W_f^0 = (m-1) {\rho v_t^2 }/{\kappa_0}
\end{equation}
Note that in this calculation we have ignored the upper limit for $\kappa$ as most of turbulent energy resides in the largest scale. The index of the turbulence, m is set to 5/3 for Kolmogorov turbulence and 3/2 for Kraichnan turbulence \citep{1965PhFl....8.1385K}. 

This fluid turbulence generates Alfv\'en waves through Lighthill radiation \citep{1952RSPSA.211..564L,1955ApJ...121..461K}. Continuing to follow \cite{1984ApJ...277..820E} and \cite{Fujita03}, we limit the scale for the largest Alfv\'en waves to correspond to the Taylor length, which defines the transition between large-scale ordered fluid motions and small-scale disordered motions:

\begin{equation}\label{eq:l_T}
l_T = l_0\,\left(\frac{15}{\rm Re}\right)^{1/2}
\end{equation}
Correspondingly, $\kappa_T=2\pi/l_T$ defines the minimum wavenumber of the Alfven mode. 
Here ${\rm Re} = l_0 v_t/\nu_\kappa$ is the Reynolds number of the ICM, where $\nu_\kappa=u_p\lambda_{\rm eff}/3$ is  the kinetic viscosity of ICM;    $m_p\,u_p^2/2 = k_B T$  is the thermal velocity of protons, where $T$ is the ICM temperature. The value $\lambda_{\rm eff} = \lambda_g^2/\lambda_c$ is the mean free path for the transverse drift of protons in cluster magnetic fields, where $\lambda_g = {u_p m_p c}/{eB}$ is the proton gyroradius and $\lambda_c= {3^{3/2}\,(k_B T)^2}/{(4\pi^{1/2}\,n_p\,e^4\,\ln \Lambda)}$ is the mean free path for Coulomb collisions \citep{1962pfig.book.....S},  where $\ln \Lambda\approx \ln(12\pi n_p \lambda_{\rm De}^3)$  is the Coulomb logarithm and $\lambda_{\rm De} = \sqrt{k_B T/n_p e^2}$ is the Debye length.  
For typical values of the above parameters we obtain a Reynolds number:
\begin{eqnarray} 
\label{eq:reynolds}
{\rm Re}&=&10^{26}\,\left(\frac{l_0}{300\,\rm kpc}\right)\left(\frac{v_t}{300\,\rm km\, s^{-1}}\right)\left(\frac{40}{\ln\Lambda}\right) \\ \nonumber
&\times& \left(\frac{n_p}{10^{-3}\,\rm cm^{-3}}\right)^{-1}\left(\frac{T}{2\,\rm keV}\right)^{1/2}\left(\frac{B}{1\,\mu\rm G}\right)^2
\end{eqnarray}

For each wavenumber $\kappa$ exceeding the threshold $\kappa_{T}$, Alfv\'en waves are generated at a wavenumber k, calculated as in \citep{Fujita03}:
\beq\label{eq:k_kappa}
k = \kappa\,\frac{v_f(\kappa)}{v_A}
\eeq
where  $v_f(\kappa)$ is the   velocity of an Eddy with wavenumber $\kappa$, which can be calculated by  $v_f\sim \left({\kappa W_f(\kappa)}/{\rho}\right)^{1/2}$. Plugging in Equations~\ref{eqn:W_f} and~\ref{eqn:Wf0} we get
\beq
v_f(\kappa) = v_t\,(m-1)^{1/2}\,\left(\frac{\kappa}{\kappa_0}\right)^{(1-m)/2}
\eeq
 and
$v_A$ is the Alfv\'en velocity in plasma with magnetic field B \citep{1942Natur.150..405A}, 
\beq
v_A=\frac{B}{(4\pi\rho)^{1/2}} =70\,\left(\frac{B}{1\mu\rm G}\right)\,\left(\frac{n_p}{10^{-3}\,\rm cm^{-3}}\right)^{-1/2}\, \rm km\,s^{-1}
\eeq

Assuming that the energy injected into Alfv\'en waves from turbulence follows a power law
\beq
I_A(k)=I_0 \left(\frac{k}{\kappa_T}\right)^{-s_t}
\eeq
where $I_A$  has the units of energy per volume per time per wave number.  Within the context of Lighthill theory, $s_t$ can be calculated as
\beq \label{eq:St}
s_t =\frac{3(m-1)}{3-m}
\eeq
which, combined with Equations~\ref{eqn:W_f} and \ref{eqn:Wf0}, gives \citep{1984ApJ...277..820E, Fujita03}:
\beq\label{eqn:I0}
I_0 = \eta_A(s_t -1)\rho v_A^3\left(\frac{v_t^2}{v_A^2 R}\right)^{3/(3-m)}
\eeq  
where $R=(m-1)^{-1}({\kappa_0}/{\kappa_T})^{1-m}$, and $\eta_A \sim |2(3m-4)/(3-m)|$ is a factor of order unity \citep{Brunetti04}. The portion of the total turbulent power going into the Alfv\'en waves can be calculated by $P_A=\int_{k_T}^{k_{\rm max}} I_Adk = \eta_A\left({v_t^2}/{v_A^2 R}\right)^2 \rho v_A^3 \kappa_T$. Assuming the same parameters as in Equation~\ref{eq:reynolds}, the power is calculated as:

\begin{eqnarray}\label{eq:P_A}
P_A&=& 4.2\times10^{-32}\,\left(\frac{v_t}{300\,\rm km\,s^{-1}}\right)^{23/6}\\ \nonumber
&\times& \left(\frac{B}{1\,\mu\rm G}\right)^{-4/3} \left(\frac{n_p}{10^{-3}\,\rm cm^{-3}}\right)^{5/3}\left(\frac{T}{2\,\rm keV}\right)^{-1/12} \\ \nonumber
&\times&\left(\frac{l_0}{300\,\rm kpc}\right)^{-7/6}\,\rm erg\,cm^{-3}\,s^{-1} 
\end{eqnarray}

As a comparison, assuming that the power injection takes $\sim 1$ Gyr, the Alfv\'en energy density is just  $0.04\%$ of the thermal energy density, $\mathcal{E}_{\rm th}=3.2\times10^{-12}\,(n_p/10^{-3}\,{\rm cm^{-3}})(T/2\,{\rm keV})\,\rm erg\,cm^{-3}$.

The spectrum of the Alfv\'en waves, $W_k(t)$, responds to this energy input, and evolves due to both wave-wave and wave-particle coupling. These processes  can be described by a continuity equation \citep{1979ApJ...230..373E}:

\beq\label{eq:Wk_continuity_complete}
\frac{\partial W_k(t)}{\partial t} = - \sum\limits_{i=1}^n \Gamma_k^i W_k(t) + I_A(k, t)
\eeq

The first term denotes the wave damping with relativistic particles in the ICM, which we will discuss next in Sec.~\ref{sec:reacc}. 
We note that in a complete description of the system, a cascade term ${\partial}\left[ D_{kk} {\partial W_k(t)}/{\partial k}\right]/{\partial k}$ should also be included on the right-hand side of Equation~\ref{eq:Wk_continuity_complete}. This physically corresponds to a wave-wave interaction term, with $D_{kk}$ being the wave-wave diffusion coefficient. In the remainder of this work, we follow \citep{Fujita03} and ignore this term in our analytic approximation. However, this approximation depends on both the energy density of the waves and the relativistic particle spectrum. We refer readers to~ \citep{Brunetti04} for a detailed comparison of the relevant time-scales over a number of physical environments.


\subsection{Alfv\'enic Acceleration}\label{sec:reacc}
Alfv\'en waves can accelerate particles via resonant damping.   In order to resonate with Alfv\'en waves,  an input population of non-thermal electrons is needed \citep{1992ApJ...398..350H}. While we do not comment specifically on the origin of this charged particle population, we note that non-thermal protons and electrons are known to populate galaxy clusters. Possible astrophysical injection mechanisms for these relativistic particles include collisionless shocks in merger events \citep{2000ApJ...535..586T,2009MNRAS.395.1333V}, active galactic nuclei  \citep{1997ApJ...477..560E} and powerful galactic winds during starburst activity \citep{1999APh....11...73V}  (We refer readers to \cite{2014IJMPD..2330007B} for a detailed review).  In our model, we simply assume that there is a sufficient density of non-thermal electrons to be re-accelerated by the turbulence powered by major mergers. 

The resonance condition between a wave with wave number $k$ and a particle with mass $m$ and momentum $p$ can be written as  \citep{1979ApJ...230..373E}

\beq\label{eq:resonanceCondition}
k = \frac{\Omega m}{p} \,\frac{1}{\mu \mp v_A/c}
\eeq
where $\Omega = eB/mc$ is the cyclotron frequency,    $\mu = v_{\parallel}/c$ is the projected particle velocity along the magnetic field, and the minus (plus) sign corresponds to electrons (protons) as damping particles. Notice that we have used the dispersion relation $\omega\approx k_\parallel v_A$  and assumed $k\approx k_\parallel$, which is reasonable for Alfv\'en waves in the ICM \citep{Brunetti04}. 

The resonance condition indicates a relation between the minimum wave number and the maximum Lorentz factor of electrons. Considering 
that $v_A\ll c$ and $|\mu|$ is of order unity, Equation~\ref{eq:resonanceCondition} implies $\gamma_{\rm max} \approx \Omega_e/k_{\rm min}=\Omega_e/k_T $.  From the definitions of Equations~\ref{eq:l_T} and~\ref{eq:k_kappa} we obtain:
 
\begin{eqnarray}
\gamma_{\rm max} = \frac{1}{(m-1)^{1/2}}\,\frac{\Omega\, v_A}{\kappa_0\, c\, v_t}\,\left(\frac{\rm Re}{15}\right)^{(m-3)/4} 
\end{eqnarray} 
 assuming consistent physical parameters with Section~\ref{subsec:turbulence}, this corresponds to a maximum electron energy: 
\begin{eqnarray}\label{eqn:Emax}
E_{\rm max} &=& 53\,\left(\frac{B}{1\mu\rm G}\right)^{4/3}\left(\frac{l_0}{300\,\rm kpc}\right)^{2/3}\left(\frac{T}{2\,\rm keV}\right)^{-1/6}\\ \nonumber
&\times& \left(\frac{v_t}{300\,\rm km \,s^{-1}}\right)^{-4/3}\left(\frac{n_p}{10^{-3}\,\rm cm^{-3}}\right)^{-1/6}\,\rm GeV
\end{eqnarray}

$E_{\rm max}$ represents the highest energy of an electron that could resonate with the turbulence, but can be considered as an estimation of the maximum energy of accelerated electrons as we will demonstrate at the end of this section. 

Using the resonance condition, the damping rate $\Gamma(k)$ in Equation~\ref{eq:Wk_continuity_complete} due to the interaction with electrons can be written as \citep{1968Ap&SS...2..171M, Fujita03}:
\beq\label{eq:Gamma}
\Gamma(k) = -\frac{4\pi^3e^2 v_A^2}{c^2 k}\int_{p_{\rm min}}^{p_{\rm max}}\left[1- \left(\frac{v_A}{c}+\frac{\Omega m_e}{pk}\right)^2\right]\,\frac{\partial f}{\partial p}p^2dp
\eeq
Following the argument in  Section~\ref{subsec:turbulence}, the wave-wave interaction term in Equation~\ref{eq:Wk_continuity_complete} can be ignored, so the electron number density and the Alfv\'en wave spectrum evolve according to the following coupled equations \citep{1984ApJ...277..820E}:
\begin{eqnarray}
\frac{\partial W_k(t)}{\partial t} &=& -  \Gamma(k) W_k(t) + I_A(k, t)  \label{eq:wave} \\
\frac{\partial f}{\partial t} &=& \frac{1}{p} \frac{\partial}{\partial p} \left[p^2D_{pp}\frac{\partial f}{\partial p} + Sp^4 f\right] \label{eq:particle}
\end{eqnarray}
where $f(p) = N(p)/(4\pi p^2) $ is the phase space electron distribution;    $Sp^2c$ corresponds to  the synchrotron and inverse Compton emission power of an electron with  $S = 4(B^2+B_{\rm CMB}^2)e^4/(9m_e^4c^6)$; $B_{\rm CMB}=3.25\,(1+z)^2\,\mu\rm G$ is the equivalent magnetic field strength of the CMB; $D_{pp}$ is the particle diffusion coefficient due to the resonant scattering of Alfv\'en waves that are assumed to be isotopically distributed \citep{1984ApJ...277..820E}:
\beq\label{eq:D_pp}
D_{pp} = \frac{2\pi^2e^2v_A^2}{c^3}\int_{k_{\rm min}}^{k_{\rm max}}\frac{W_k}{k}\left[1-\left(\frac{v_A}{c}+\frac{\Omega m_e}{pk}\right)^2\right]dk
\eeq
Numerical solutions for the set of coupled differential equations given in Equations~\ref{eq:wave} and \ref{eq:particle} have been presented in e.g., \cite{Brunetti04} and \cite{Brunetti05}.  Here, however, aiming to study the cumulative effect of turbulent acceleration of cluster mergers, we simplify the calculation of individual events by using an analytical steady-state solution following \cite{1984ApJ...277..820E} and \cite{Fujita03}.

In the steady-state regime, we further simplify Equations~\ref{eq:wave} and~\ref{eq:particle} by setting the condition $dW_k/dt=0$ and  assuming that the wave and particle spectra are simple power-laws:
\begin{eqnarray}
W_k = W_k^0 \,k^{-\nu} \\
f(p) = f_0 \,p^{-s}
\end{eqnarray}
Noting that the lower limits of Equations~\ref{eq:Gamma} and \ref{eq:D_pp} are given by $p_{\rm min}= \Omega m_e/k$ and $k_{\rm min} = \Omega m_e/p$ respectively, the power-law assumptions lead to:
\begin{eqnarray}
\Gamma(k) &=& \frac{2}{s-2}\frac{4\pi^3e^2v_A^2f_0}{kc^2}\left(\frac{\Omega m_e}{k}\right)^{2-s}\\
\label{eqn:Dpp}
D_{pp}&=& \frac{(s-2)\kappa_T^{s_t}}{2\pi\nu(\nu+2)c}\frac{I_0}{f_0}\left(\Omega m_e\right)^{s-2-\nu}p^\nu\equiv Ap^\nu \\ 
s &=& \nu+3 - s_t 
\end{eqnarray} 
As demonstrated in \cite{1984ApJ...277..820E, Fujita03},  if and only if 
\beq
\nu=3
\eeq
and 
\beq\label{eqn:SA}
S=sA\, ,
\eeq
a  self-similar solution with $\partial f/\partial t = 0$ can be reached :
\beq\label{eq:f_p}
f(p)\propto p^{-(6-s_t)}
\eeq 
With this spectrum and using Equation~\ref{eq:P_A}, the energy loss rate of electrons is found to satisfy the relation
\begin{eqnarray}\label{eq:P_SI}
P_{\rm SI} &\equiv& \int Sp^2c f(p) 4\pi p^2dp \\ \nonumber
&=& P_A\,\frac{2}{15}(6-s_t)(4-s_t)
\end{eqnarray}
Taking $m=5/3$ for Kolmogorov turbulence and using Equations~\ref{eq:St}, \ref{eq:f_p}, \ref{eq:P_SI},  as well as the $\eta_A$ factor below equation~\ref{eqn:I0}, we find a synchrotron and inverse Compton emission power that is comparable to the energy flux injected in the Alf\'ven waves $P_{\rm SI}=P_A$,  and  an electron energy spectrum $N(E)\propto E^{-2.5}$. Specifically, Equation~\ref{eq:P_SI} normalizes the electron flux through the relation $P_{\rm A}\approx (16\pi/3) S f_0\,p_{\rm max}^{1/2}$, and thus the equilibrium   spectrum can be written as:
\beq\label{eqn:dNdEe}
\left(\frac{dN_e}{dE_edV} \right)_{\rm inj}= \frac{3P_A\,c}{4 S(E_{\rm max})^{1/2}}\,E_e^{-5/2}
\eeq

\begin{figure*}
\centering
\epsfig{file=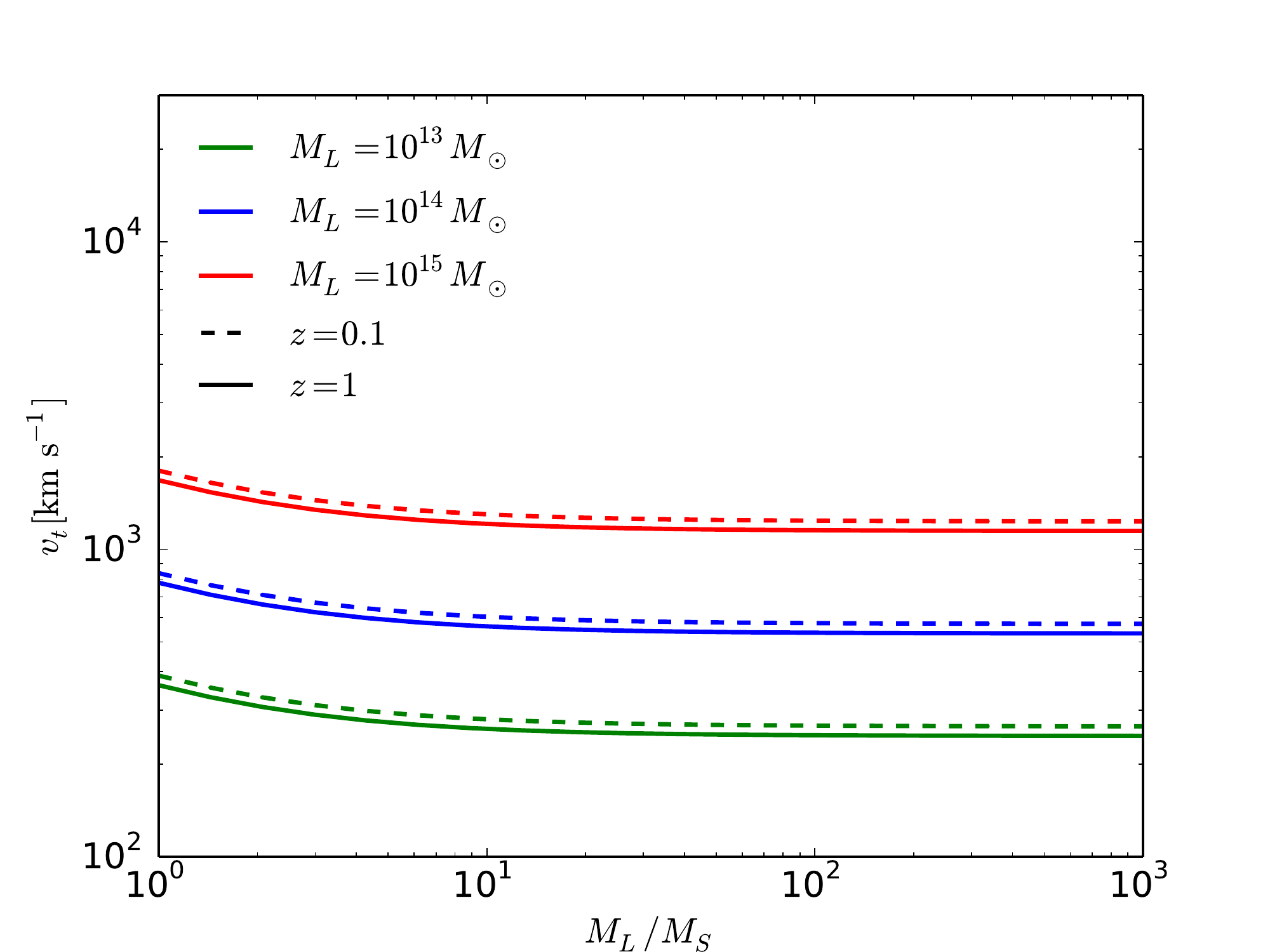,width=0.45\textwidth}   
\epsfig{file=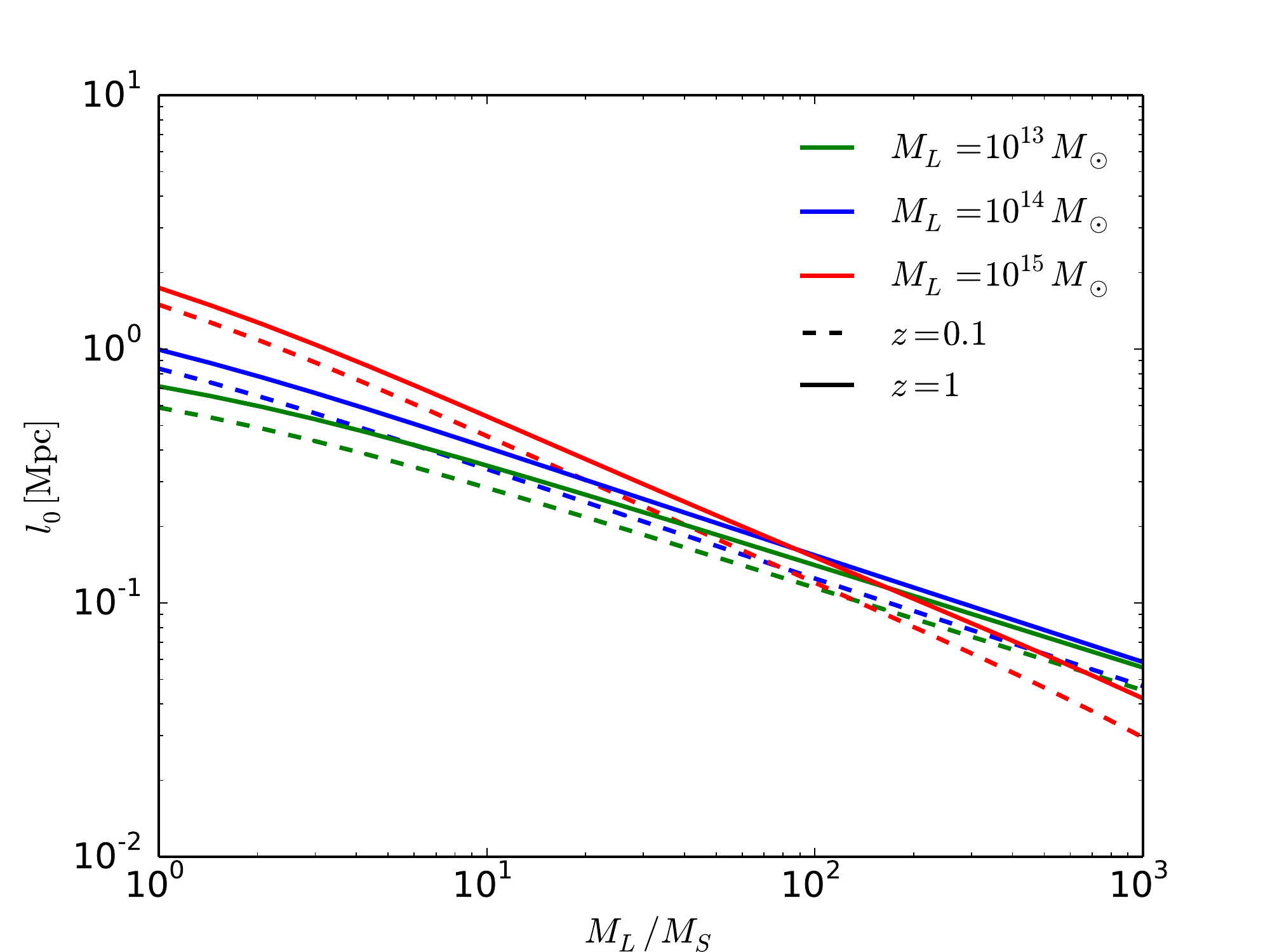,width=0.45\textwidth}   
\epsfig{file=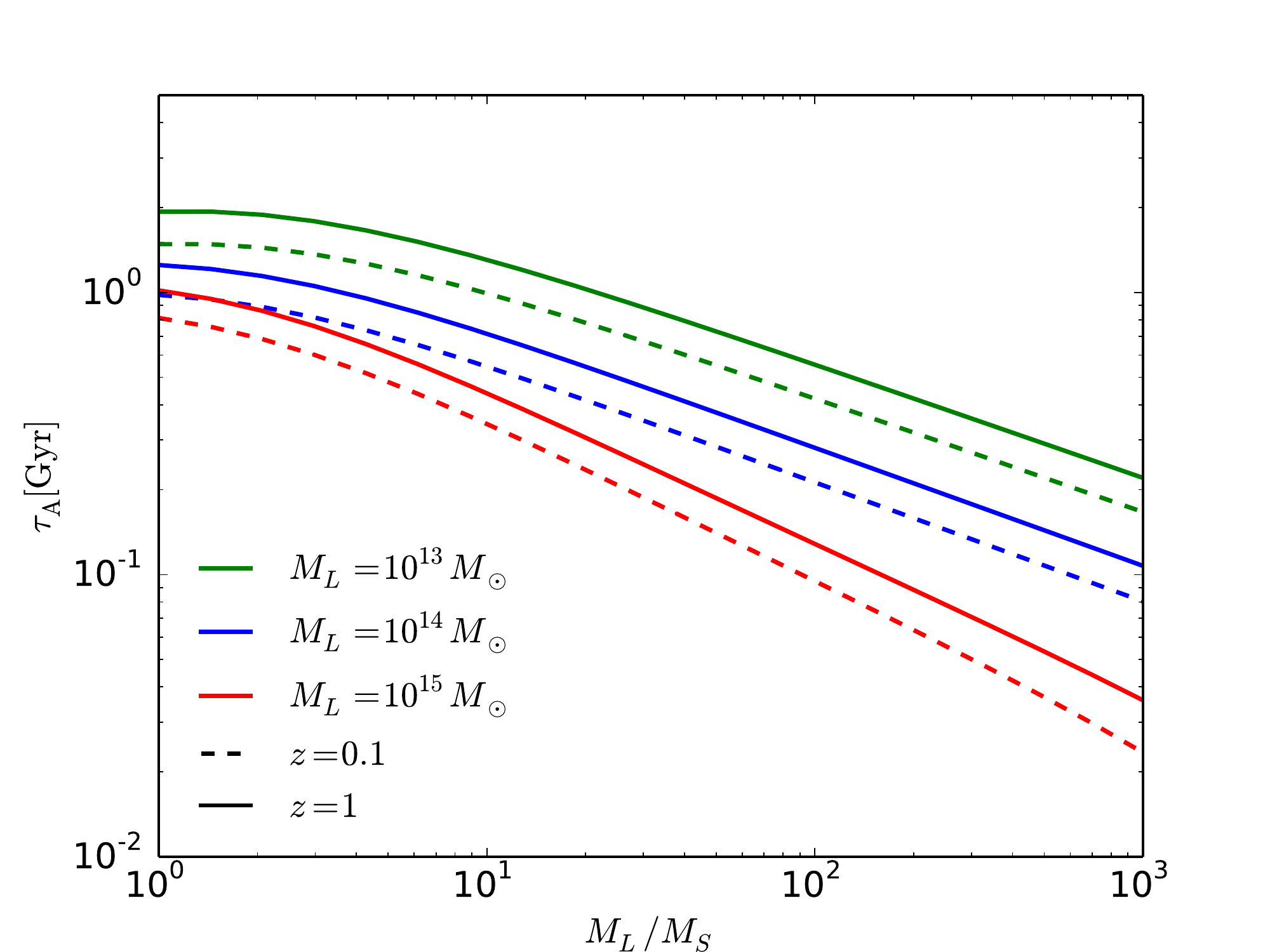,width=0.45\textwidth}   
\epsfig{file=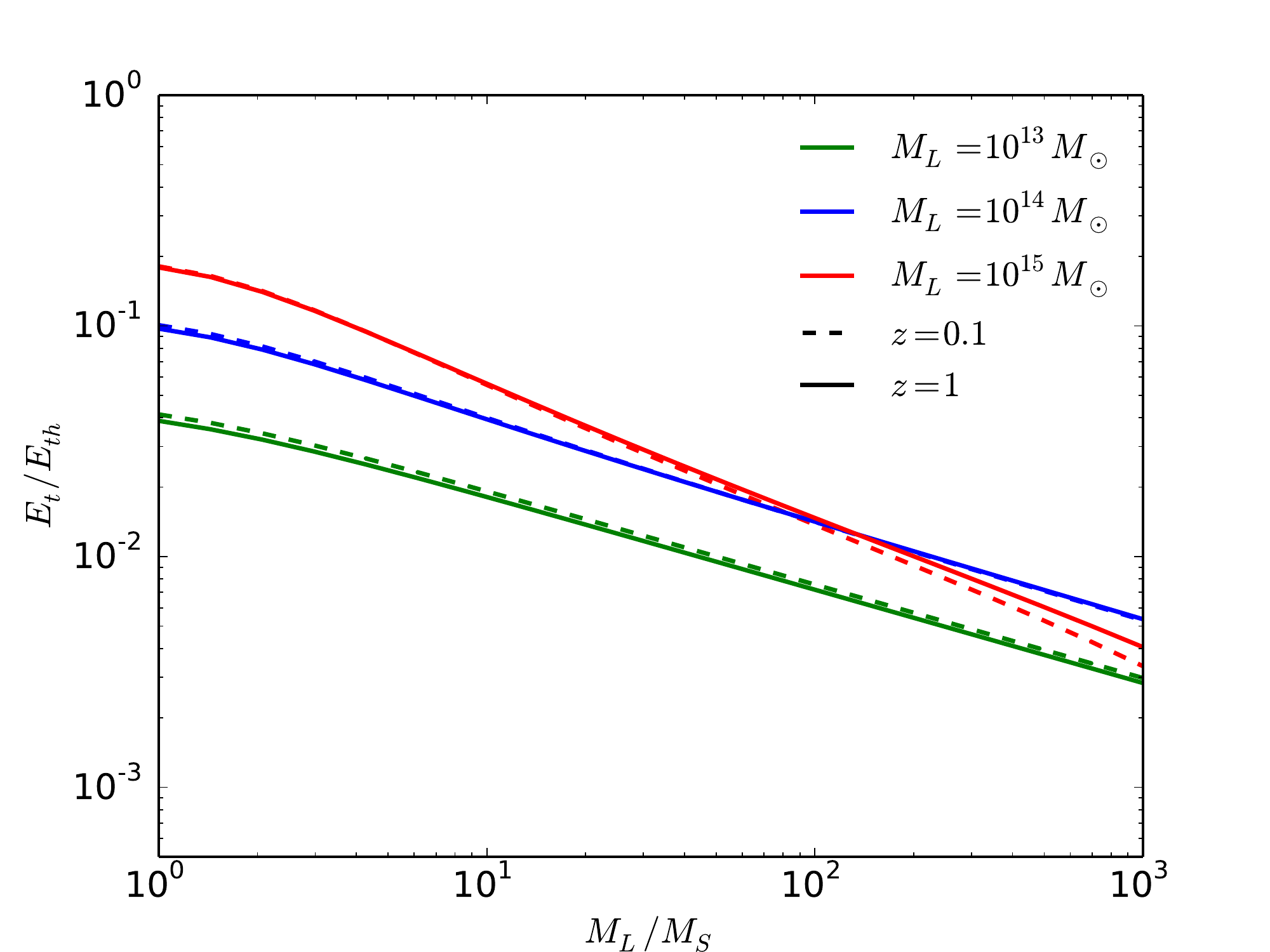,width=0.45\textwidth}   
\caption{\label{fig:turb_properties} Properties of the turbulence generated during the merger of two halos as a function of the halo mass ratio.  Top panels:  Turbulent velocity at injection (left) and the injection scale of the turbulent Eddies (right). Bottom panels:   Timescale of the turbulence $\tau_A=l_0/v_t$ (left),  Ratio between the turbulent energy and the thermal energy of  the host halo  (right).   Values are evaluated at two redshifts: $z=0.1$ (dashed lines) and $z=1$ (solid lines), and for three different masses of the host halo: $M_L=10^{15}\,M_\odot$ (red), $10^{14}\,M_\odot$ (blue), and $10^{13}\,M_\odot$ (green). $\eta_v=0.8$ has been assumed for all calculations.}
\end{figure*}

\noindent The spectrum of synchrotron emission is given by:
\begin{equation}
\frac{dN_s}{dE_sdtdV} = \frac{\sqrt{3}e^3B}{hm_e c^2\,E_s}\,\int_{m_e}^{E_{\rm max}}dE_e\,F(\frac{E_s}{h\nu_c})\,\frac{dN_e}{dE_edV}
\end{equation}
where $F(x)=x\int_x^\infty K_{5/3}(x')dx'$ is the Synchrotron function, $K_{5/3}$ is the modified Bessel function, and $\nu_c$ is the critical frequency of synchrotron emission, $\nu_c =  {3}\gamma^2{eB}/{(4\pi m_e c )} = 1.6\, ( {B}/{1 \mu \rm G})({E_e}/{10 \rm GeV})^2 ~ \rm GHz $. Note that the synchrotron emission function $F(x)$ peaks at $x = \nu/\nu_c\sim~0.29$.

While  $E_{\max}$ in Equation~\ref{eqn:Emax} represents the maximum energy an electron can gain through damping, our self-similar solution is strictly time-independent. This indicates that the acceleration timescale for electrons $t_{\rm acc}\sim p^2 / 2 D_{pp}$ is equal to the energy loss timescale $t_{\rm loss} \sim  {E}/({dE/dt}) = c\,/ (SE)$  at all energies. Thus, from the analytic formula derived here it is difficult to directly compute the maximum electron energy after energy losses are taken into account. However,  the relevant timescales have been computed   in  \citep{Brunetti04}, who find that it is possible to accelerate electrons to at least $\gamma\sim10^{4.6}$, and possibly much higher as the calculations of \cite{Brunetti04} scan a limited range of magnetic field strengths and time-ranges. 
While additional magnetohydrodynamic studies are necessary to fully explore these models in the context of the ARCADE excess, we note that our results are not highly sensitive to the maximum electron energy. In Figure~\ref{fig:MHD_compare} of the appendix, we will show that the fit to the ARCADE data is only mildly impacted at high-frequencies when the maximum electron energy is decreased. In part, this is due to the fact that a significant fraction of the total synchrotron signal is produced by clusters with intermediate masses, as we will show in Figure~\ref{fig:yT_m}), which have a corresponding $\gamma_{\rm max}$ much smaller than $10^5$. Finally, we note that Alfv{\'e}n reacceleration can also produce high-energy electrons with energy $\gamma$~$>10^{5}$ through a second process. Protons accelerate to very high energy via Alfv{\'e}n reacceleration can produce high-energy electrons through inelastic collisions with the interstellar medium \citep{Brunetti05}. This process potentially provides a new source of high-energy electrons  capable of increasing the synchrotron emission above $\sim$5~GHz.

We stress that calculations in the above two sections are entirely based on the Eilek-Henriksen solution \cite{1984ApJ...277..820E}, which is strictly true only under ideal circumstances, when waves mainly interact with electrons  (but not protons),  wave-electron interaction dominates over wave-wave cascading, and   energy losses follow $p^2$ scaling (like inverse-Compton and synchrotron processes).  A magnetohydrodynamic approach like in \cite{Brunetti04, Brunetti05} would be needed to fully describe the system. Yet the scenarios discussed in \cite{Brunetti04, Brunetti05} could be further complicated by the fact that Alfv\'en waves can develop a
scale-dependent anisotropy during their cascading (e.g., \cite{1995ApJ...438..763G, 2003ApJ...595..812C, 2006ApJ...645L..25L}), which could strongly reduce the acceleration rate  \cite{2000PhRvL..85.4656C}; while on the other hand,   compressive turbulence in the
ICM could enhance the acceleration (for example fast modes, see \cite{2007MNRAS.378..245B,2015ApJ...800...60M}). The current work, however, aims to explore the possible link between the ARCADE-2 excess and  reacceleration of electrons induced by cluster mergers. Therefore we limit our calculation of the electron spectrum to the ideal case.

\subsection{Density Profile, Magnetic Field and Temperature of a Cluster}

We assume a simple $\beta$-model for the  density profile of the thermal gas in the ICM   \citep{1976A&A....49..137C, Brunetti05}:
\beq
\rho_{\rm ICM}(r) = \rho_{\rm ICM, 0}\,\left[1+\left(\frac{r}{r_c}\right)^2\right]^{-3\beta/2}
\eeq
where $\beta=0.8$ and $r_c$ is the core radius which is taken as 10\% of the cluster's virial radius $ r_{\rm vir}$ \citep{Fujita03}. 
The virial radius is determined through the relation $M~=~4\pi r_{\mathrm {vir}}^3\triangle_{\mathrm {vir}}(z)\,\rho_m(0)/3$, with $\triangle_{\mathrm {vir}} (z) \approx (18\pi^2 + 82 x - 39x^2)/(x+1)$ and $x =\Omega_M(z) - 1$ \citep{2003ApJ...584..702H}.
The density profile is normalized by: 
\beq
f_{\rm ICM}  M = \int \rho_{\rm ICM}(r) 4\pi r^2dr 
\eeq
with $f_{\rm ICM}\sim 0.13 (M/ 10^{14}M_\odot)^{0.16}$ the baryon fraction of clusters \citep{2013ApJ...778...14G}.

For clusters, which we define as any halo with a total mass $M\ge10^{12}M_\odot$, we adopt a magnetic field distribution as a function of the radial distance from the center of the cluster to be \citep{Dolag:2001vy, Donnert:2008sn,2010A&A...513A..30B}
\label{subsec:bfield}
\begin{equation}\label{eqn:B_cluster}
B(M,r) = B_0  \left(\frac{M}{M_*}\right)^{\alpha} \left[1 + \left(\frac{r}{r_{\mathrm{c}}}\right)^2\right]^{-3\beta \eta/2}
\end{equation}
where $\eta\sim 3/8$ \citep{FL14}, and $B_0$ is the central magnetic field strength of clusters. We leave $B_0$ as our first adjustable parameter with a default value B$_0$~=~10$\mu$G.  
The central magnetic field strengths  should also depend on the  temperature, and therefore the mass of clusters as suggested in  e.g. \cite{Govoni:2004as, 2005JCAP...01..009D}.  
 Here we assume that the field strength scales to the mass of cluster by $(M/M_*)^\alpha$, with $M_*=10^{14}\,M_\odot$ and $\alpha\sim0.3$ (inferred from the estimations of field strengths of Coma and A3667 in \cite{Govoni:2004as}). 
For galaxy-sized halos with magnetic fields dominated by their dense baryonic cores, we assume a  magnetic field model \citep{hooper_arcade_excess}
\begin{equation}\label{eqn:B_galaxy}
B(r) =\max\left( B_1 \,e^{-r/R_1},  B_2\,e^{-r/R_2}\right)
\end{equation}
where B$_1$ = 7.6 $\mu$G and R$_1$ = 0.025~r$_{\mathrm{vir}}$ and B$_2$~=~35~$\mu$G with R$_2$ = 0.008r$_{\mathrm{vir}}$. 

As we will show in Section~\ref{sec:results}, the cumulative radio signal   is dominated by clusters with mass greater than $10^{14}\,M_\odot$. Therefore our result is mostly independent on the magnetic field models for galaxy-sized halos. As for the magnetic field model for clusters, our default assumption of $B_0=10\,\mu$G is close to the upper limit of cluster magnetic field strength derived from  Faraday rotation measurements \citep{2002ARA&A..40..319C}. Nevertheless,  a significant enhancement of the magnetic field would be expected  due to the boost in the gas density by the shocks during merger events \citep{2002ASSL..272.....F,2012SSRv..166..187B}. For comparison, our default model predicts $B_{\rm vir}=2.5\,\mu$G at the virial radius of a $10^{15}\,M_\odot$ halo during the merger, while observations of radio relics suggest $\mu$G-level magnetic field at the position of the shock fronts \citep{2012A&ARv..20...54F}.

The temperature of ICM gas can be described by \citep{Fujita03}
\beq
T_{\rm ICM} = T_{\rm vir}  + \frac{3}{2}T_g
\eeq
where $3T_g/2=0.8\,\rm keV$ is the temperature of the preshock gas   \citep{Fujita03}  and $T_{\rm vir}$ is the virial temperature of a cluster, \beq
T_{\rm vir} = \frac{GM\,\mu_p m_p}{2\,r_{\rm vir} \,k_B}
\eeq
with $\mu_p=0.61$ being the mean molecular weight  and $k_B$ being the Boltzman constant. 

\subsection{Cluster Mergers}

Turbulence is expected to be injected during the passage of subhalos \citep{2005MNRAS.357.1313C}. 
In what follows, we denote the smaller (sub-) halo with a subscript 'S', and the larger (host) halo with the subscript 'L'. 
The relative velocity of the two halos with masses of $M_L$ and $M_S$ can be described by \citep{2002ASSL..272....1S}
\beq
v_i =\left[\frac{2G(M_L + M_S)}{R_L}\left(1-\frac{1}{\eta_d}\right)\right]^{1/2}
\eeq
with $\eta_d = 4\,\left((M_L + M_S)/M_L\right)^{1/3}$. 
Following \cite{Fujita03} and \cite{2005MNRAS.357.1313C}, we estimate the turbulent velocity by 
\beq
v_t = \eta_v\, v_i
\eeq
with $0<\eta_v<1$   being  our second free parameter.  The top left panel of Figure~\ref{fig:turb_properties} shows $v_t$ as a function of the halo mass ratio $M_L/M_S$, 
 taking $\eta_v=0.8$. 
The turbulent velocity at the injection scale is around $300-2000\,\rm km\,s^{-1}$ depending on the mass of the host halo, a value that is consistent with previous studies   (e.g. \citep{2006MNRAS.366.1437S}). Note that when $M_L/M_S\gg1$, the turbulent velocity is approximately the free-fall velocity due to the gravitational potential of $M_L$, yielding $v_t\sim GM_L/R_L$. 

\begin{figure}
\centering
\epsfig{file=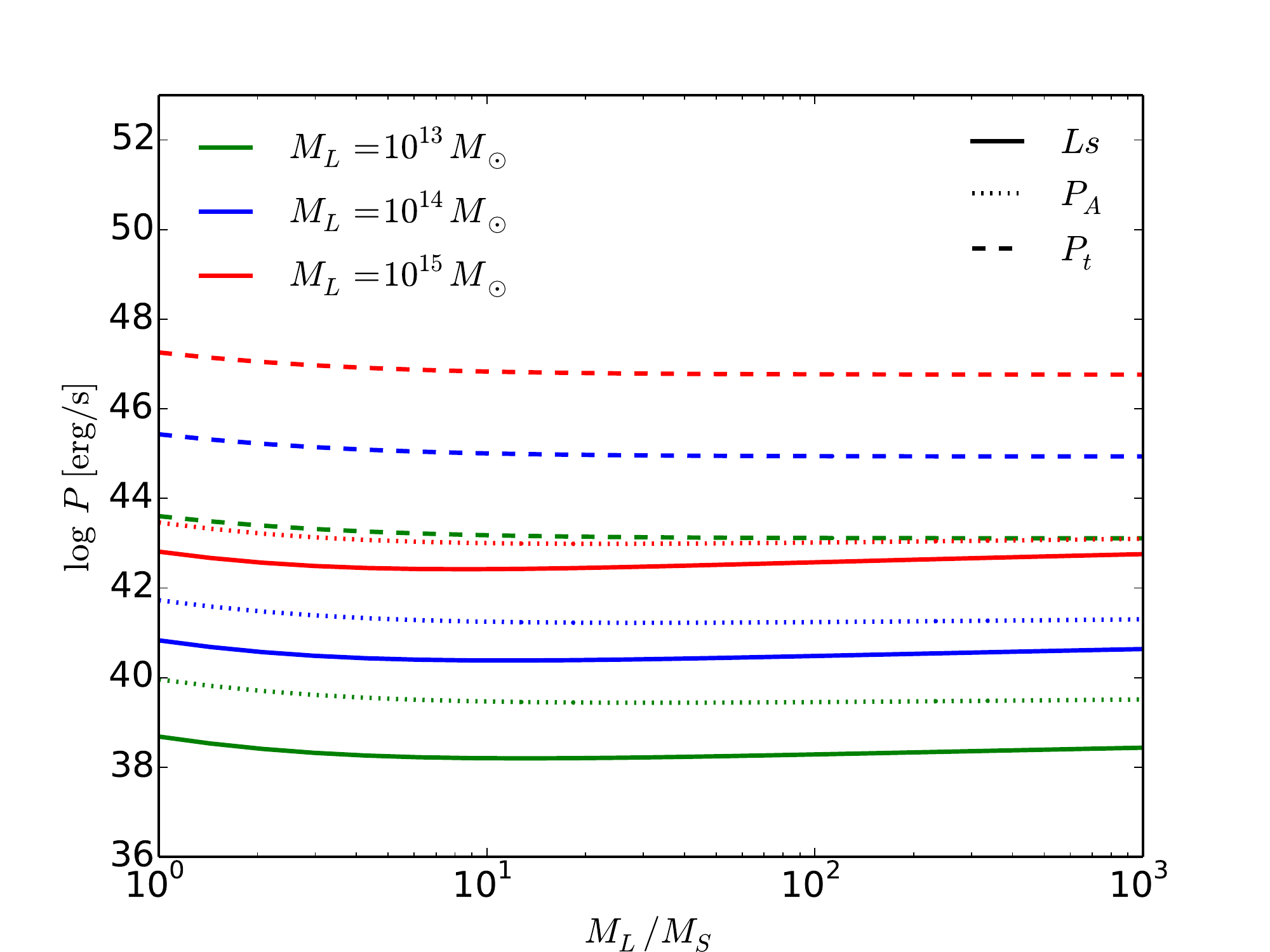,width=0.7\textwidth}   
\caption{\label{fig:power} Energy power of the injected turbulence (dashed), the Alfven waves generated by the turbulence (dash-dotted), and the synchrotron emission from electrons accelerated by the Alfv\'en waves (solid). As in Figure~\ref{fig:turb_properties}, values are calculated for three host halo masses, $M_L=10^{13}, 10^{14}, 10^{15}\,M_\odot$ in green, blue and red, respectively. Parameters assumed in this plot include central magnetic field strength   $B_0=10\,\mu$G and  $\eta_v=0.8$.   }
\end{figure}

The smaller, infalling halo suffers ram-pressure stripping  when it crosses the larger one. At the stripping radius, the ram pressure applied to the smaller halo  is comparable to its static pressure:
\beq
\bar{\rho}_L v_i^2=\frac{\rho_S(r_s)k_B T_S}{\mu m_p}
\eeq
Turbulence is expected to be  injected on a similar scale as the stripping radius. In \cite{Fujita03} and \cite{Brunetti04},  $l_0=\eta_l \,r_s$ was assumed, taking $\eta_l$ to be another free parameter.
However, noting that $\eta_l$ is poorly constraint observationally, and that the Alfv\'en power is significantly less dependent on $l_0$ than on $v_t$ (recalling that $P_A\propto v_t^{23/6}\,l_0^{-7/6}$), in this work we reduce the degrees of freedom in the model and set  $l_0=2\,r_s$. 
We show $l_0$ as a function of the halo mass ratio in the top right panel of Figure~\ref{fig:turb_properties}. Note that in the extreme case   $M_L\gg M_S$, $r_s\rightarrow 0$, meaning that the smaller halo is almost immediately stripped when it approaches the larger cluster.

The duration for the injection of turbulence is of the order of the crossing time, $\tau_A = {l_0}/{v_t}$ \citep{Fujita03}. As plotted  in the bottom left panel of Figure~\ref{fig:turb_properties}, the typical timescale for turbulence ranges from 0.1 to a few Gyrs depending on the ratio of the halo masses. Note that $\tau_A$ is longer for smaller host halos, due to their significantly slower $v_t$.  
The volume swept by the passage of the subhalos would be filled with turbulence. High-resolution hydrodynamical  simulations  found that in major mergers, mergers  dominate turbulent diffusion in the entire virial volume; in minor mergers, 
several effects triggered by the merger including sloshing, rotational large-scale motions and hydrodynamical instabilities would introduce turbulence to a size close to the scale radius of the subcluster \citep{2011A&A...529A..17V, 2012A&A...544A.103V}. 
We therefore assume that during $\tau_A$,  Alfv\'en waves can be excited  within a swept volume  $V_A\sim \pi R_{\rm L}^2 l_0$.
Then in the bottom right panel of Figure~\ref{fig:turb_properties} we plot the ratio between the turbulent energy $E_t \equiv \bar{\rho}_L\,v_t^2\,V_A$ and the thermal energy of the two halos $E_{\rm th}=(M_L+M_S)\, k_B\,T_L/ \mu \,m_p$. For $10^{14}-10^{15}\,M_\odot$ halos this ratio is found to be $10-18\%$, consistent with  numerical simulations \citep{2015ASSL..407..557B} as well as observations of the Coma cluster, which suggested a lower limit of $10\%$ \citep{2004A&A...426..387S}.

As a consistency check, Figure~\ref{fig:power} compares the  power contained in the turbulence injected by the merger with the Alfv\'en waves generated by the turbulence, and the synchrotron emission from electrons accelerated by the Alfv\'en waves. Note that the power is evaluated over the turbulent injection time, and is thus slightly higher for a larger halo mass ratio which has a shorter $\tau_A$.

\begin{figure}
\centering
\epsfig{file=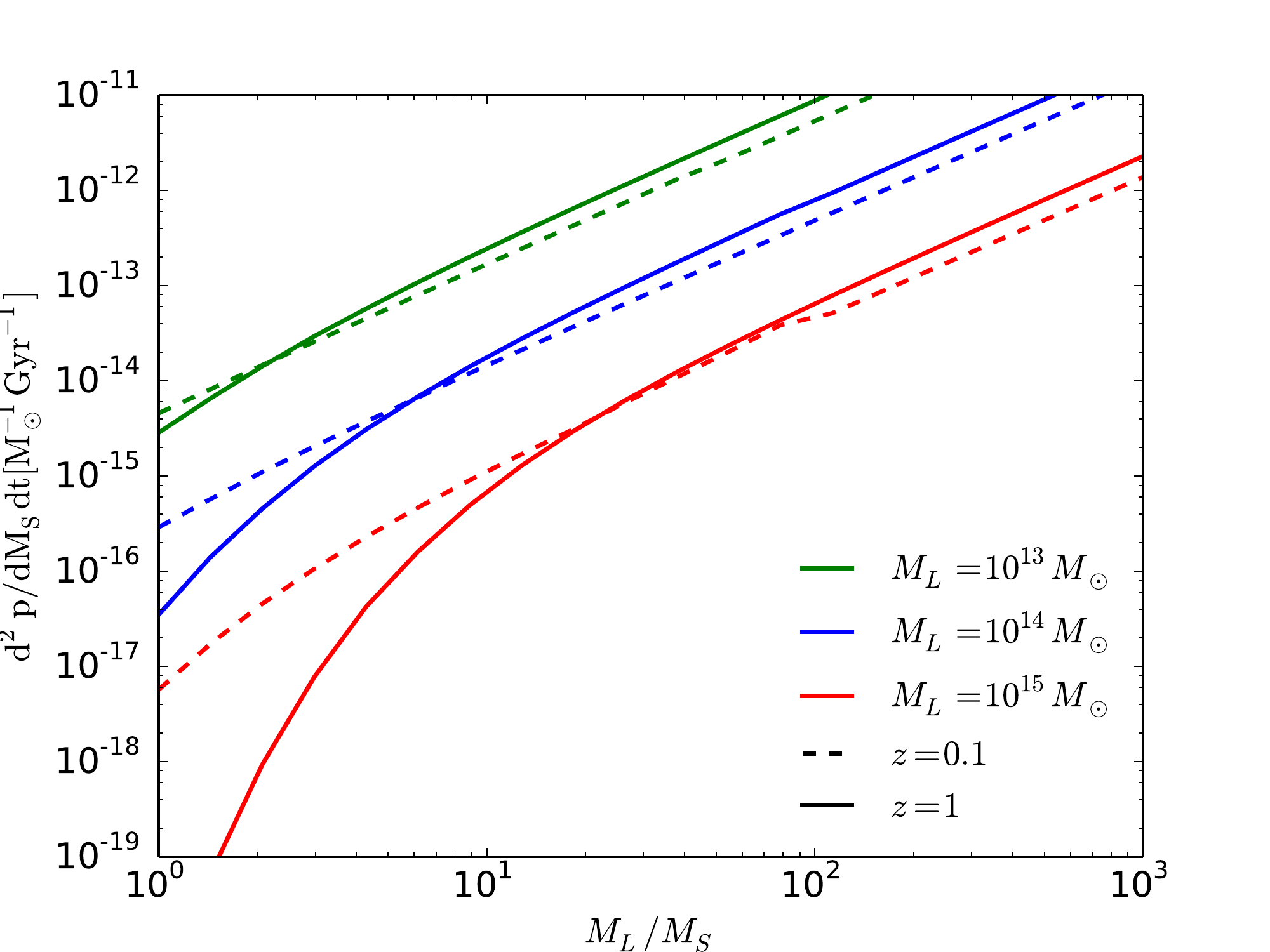,width=0.7\textwidth}   
\caption{\label{fig:mergerRate} The merger rate for a host halo of mass $M_{\rm L}$ with a smaller halo $M_{\rm S}$, as a function of the halo mass ratio $M_L/M_S$ in the comoving frame. Legends are the same as in Figure~\ref{fig:turb_properties}. }
\end{figure}

The probability that the larger cluster $M_{\rm L}$ undergoes a merger with a smaller cluster with mass $M_{\rm S}$ per unit time is given by \citep{2002ASSL..272....1S}
\begin{eqnarray}
\frac{d^2p}{dM_{\rm S}dt} &=& \sqrt{\frac{2}{\pi}}\frac{\delta_c(z)}{\sigma(M')} \left|\frac{d\ln\delta_c(z)}{dt}\right| \left|\frac{d\ln\sigma(M')}{dM'}\right| \\ \nonumber
&\times& \left[1-\frac{\sigma^2(M')}{\sigma^2(M_{\rm L})}\right]^{-3/2}\\ \nonumber 
&\times&  \exp\left[-\frac{\delta_c^2(z)}{2}\left(\frac{1}{\sigma^2(M')} - \frac{1}{\sigma^2(M_{\rm L})}\right)\right] 
\end{eqnarray}
where $M'=M_{\rm L} + M_{\rm S}$, $\delta_c(z)$ is the critical overdensity given by the expression:
\beq
\delta_c(z) = \frac{3(12\pi)^{2/3}(1+z)}{20}\left[1+0.0123\log\left(\frac{\Omega_m(z)}{\Omega_m(z)+\Omega_\Lambda}\right)\right]
\eeq
for a flat universe, $\Omega_m + \Omega_\Lambda=1$, and $\sigma(M, z)$   is the rms variance of the linear density field,
\begin{equation}
\label{eq:sigma}
\sigma^2 = \int  dk\, P^{\mathrm {lin}}(k) \tilde W(kR)\,k^2
\end{equation}
where $\tilde W(kR)$  is the Fourier transform of the real-space top-hat window function of radius $R=(3M/4\pi\rho_m)^{1/3}$ \citep{2008ApJ...688..709T} with $\rho_m(z) =\rho_m(0)\,(1+z)^3$,  and P$^{\mathrm {lin}}$ is the linear matter power spectrum   \citep{1999ApJ...511....5E}. The merger rates of three host halo masses $M_L=10^{13}, 10^{14}, 10^{15}\,M_\odot$  at $z=0.1, 1$ are shown in Figure~\ref{fig:mergerRate}. 

\subsection{Cumulative Synchrotron Intensity}

\begin{figure}
\centering
\epsfig{file=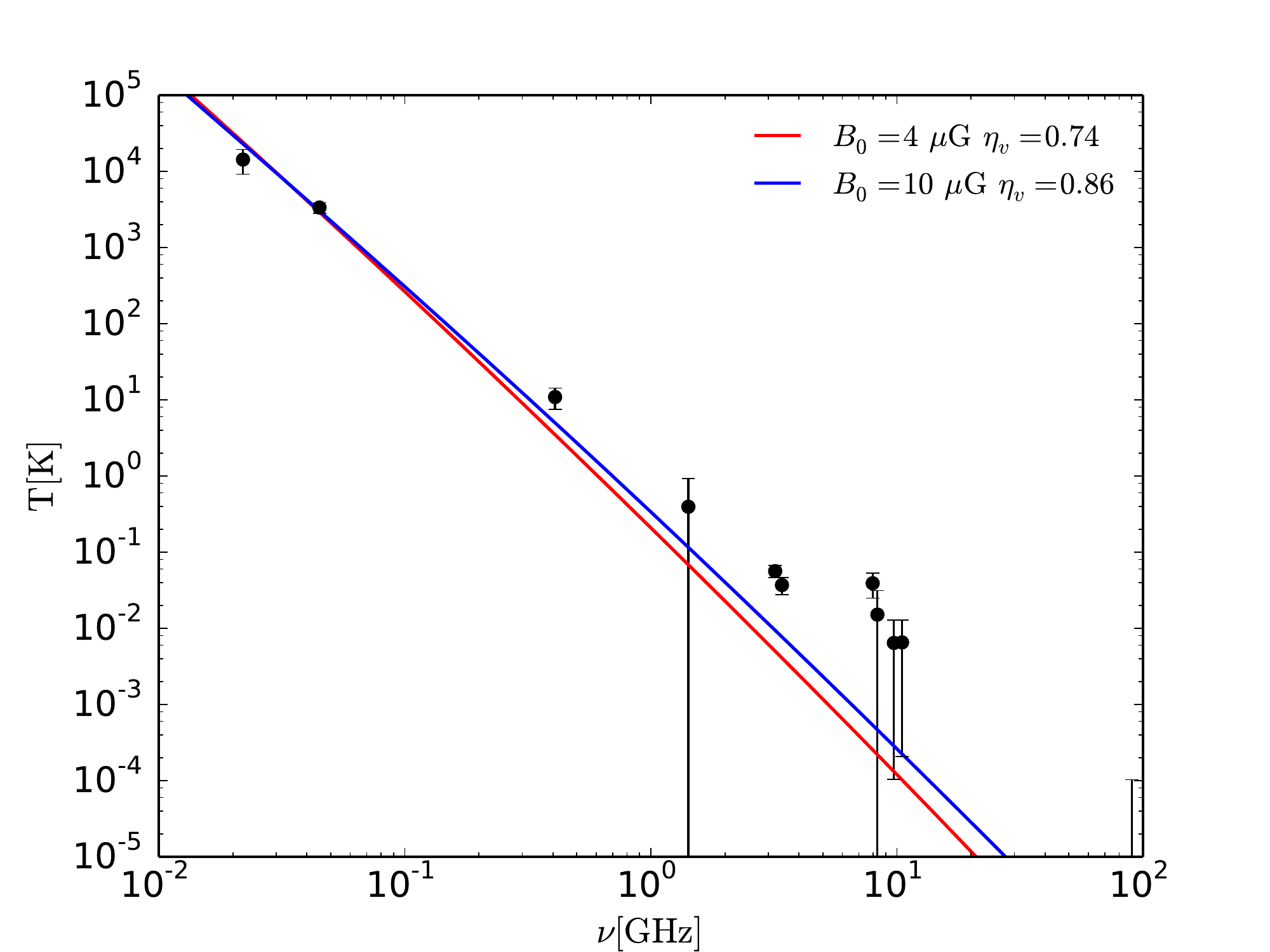,width=0.7\textwidth}   
\caption{\label{fig:yT_B} The contribution to the isotropic radio background from electrons accelerated by cluster turbulence. The synchrotron emission is shown to be comparable to the temperature excess reported by the ARCADE-2 collaboration \citep{arcade_interpretation}. Results with three sets of parameters are  presented in blue, red and green, corresponding to central magnetic field strength $B_0 = 10\,,4\,\mu$G, and a ratio between the turbulent velocity  and the initial relative velocity of the merging clusters being $\eta_v=0.86, 0.74$. }
\end{figure}

The mean intensity of synchrotron emission  from the turbulent acceleration in merging halos  can be written as
\begin{equation}\label{eqn:intensity}
I(E_{\mathrm{s}}) = \int d\chi \int_{M_{\rm L,min}}^{\infty} dM_{\rm L} \frac{dn}{dM_{\rm L}}\,W[M_{\rm L}, (1+z)E_s, z]
\end{equation}
where $\chi$ is the commoving distance, $dn/dM$ is the halo mass distribution function which is given by
\begin{equation}
\label{eq:dndM}
\frac{dn}{dM} = f(\sigma) \frac{\rho_m}{M}\frac{d \ln~\sigma^{-1}}{dM}
\end{equation} 
and for $f(\sigma)$ we use the mass function multiplicity described by  \cite{sheth_tormen_mass_function}.

\begin{figure}
\centering
\epsfig{file=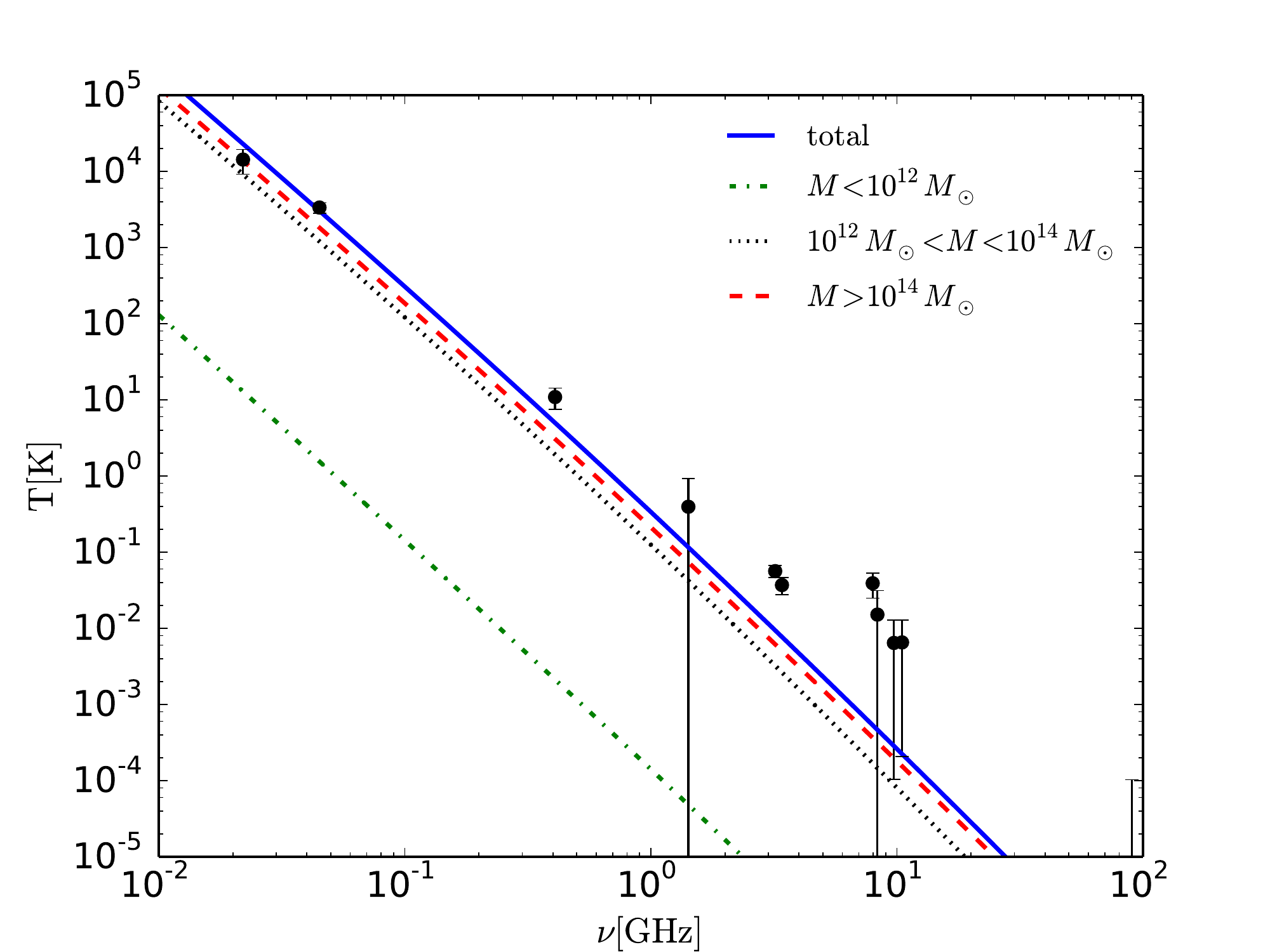,width=0.7\textwidth}   
\caption{\label{fig:yT_m} The sub-contributions to the total ARCADE-2 temperature stemming from host halos of various masses. The magnetic field  is assumed to   have a central value $B_0 = 10\,\mu$G. The turbulent velocity is assumed to be a fraction $\eta_v=0.86$  of the initial relative velocity of the merging clusters. In this figure we  show the contributions from three mass groups: $M<10^{12}\,M_\odot$, $10^{12}\,M_\odot<M<10^{14}\,M_\odot$, and $M>10^{14}\,M_\odot$. The $M>10^{14}\,M_\odot$ group (in red) dominates the signal (lying almost directly on top of the solid blue line). }
\end{figure}

The window function $W(M_{\rm L}, E_s)$ describes the synchrotron flux expected from all the merger events  of a halo $M_{\rm L}$ at redshift $z$,
\begin{eqnarray}
W(M_{\rm L}, E_s, z) &=& \frac{1+z}{4\pi}\int_{M_{\rm S,min}}^\infty dM_{\rm S}\frac{d^2p}{dM_{\rm S}dt} \\ \nonumber
&\times&  \frac{dN_s}{dE_sdt} \,\tau_A(M_{\rm L}, M_{\rm S}, z) 
\end{eqnarray}
where $dN_s/dE_sdt\,\tau_A$ sums the synchrotron emission from all radii of a larger halo during its merger with a smaller halo, with
\beq 
\frac{dN_s}{dE_sdt} = \int_0^{r_{\rm vir}} dr \,4\pi r^2 \frac{dN_s}{dE_sdVdt}(B, n_p, T, l_0, v_t)
\eeq

In the calculation we set the lower limit of $M_{\rm L}$ to be $M_{\rm L, min}=10^{11}\, M_\odot$.  This lower limit is however irrelevant, as we will show  in Sec~\ref{sec:results} that the diffusive emission is dominated by massive clusters with $M\ge 10^{14}\,M_\odot$. The lower limit of $M_{\rm S}$ is rather important, because for the same halo, its merger rate with smaller halos is significantly higher than with   larger halos, as shown in Figure~\ref{fig:mergerRate}. Considering that most of the signal comes from clusters with $M\ge 10^{14}\,M_\odot$, and that  the stripping radius of a $10^6\,M_\odot$ halo merging into a $10^{14}\,M_\odot$ halo is already close to 0,  we set $M_{\rm S,min}=10^6\,M_\odot$.

\section{Results}
\label{sec:results}

In Figure~\ref{fig:yT_B}, we present the extragalactic radio background produced from electrons accelerated by turbulence in galaxy clusters as a function of frequency. As described in Section~\ref{sec:model}, our model has two adjustable parameters: (1) the magnetic field strength  at the center of galaxy clusters (defined as halos with mass above $10^{12}\,M_\odot$), $B_0$, and (2) the ratio between the turbulent velocity  and the initial relative velocity of the merging clusters, $\eta_v$. Results with two sets of  parameters are shown in Figure~\ref{fig:yT_B}, including $B_0= 10\,,4\,\mu$G and  $\eta_v=0.86, 0.74$. 
In particular, the values of $\eta_v$ have been chosen to fit the data point at $0.05\,\rm GHz$. We find that ($B_0=10\,\mu$G, $\eta_v=0.86$) gives a better fit to the ARCADE-2 excess, while the other case with weaker central magnetic field provides a even softer spectra.

The softer synchrotron spectrum in models with lower $B_0$ is due to the constraint that the magnetic field strength at the virial radius of galaxy clusters is proportional to the central value (as given by Equation~\ref{eqn:B_cluster}). Weaker magnetic field channel a larger fraction of the turbulent energy into synchrotron emission at lower frequencies. For comparison, in a cluster with $M=10^{14}\,M_\odot$ at $z=0.1$, the magnetic field strength at the virial radius would be $B_{\rm vir}=$1.2$\mu$G for $B_0=10\,\mu$G, corresponding to a synchrotron peak frequency $\nu_p=2$ GHz for a 10 GeV electron, but would be $B_{\rm vir}=$0.5 $\mu$G and $\nu_p=$0.8 GHz  for $B_0=4\,\mu$G.

\begin{figure}
\centering
\epsfig{file=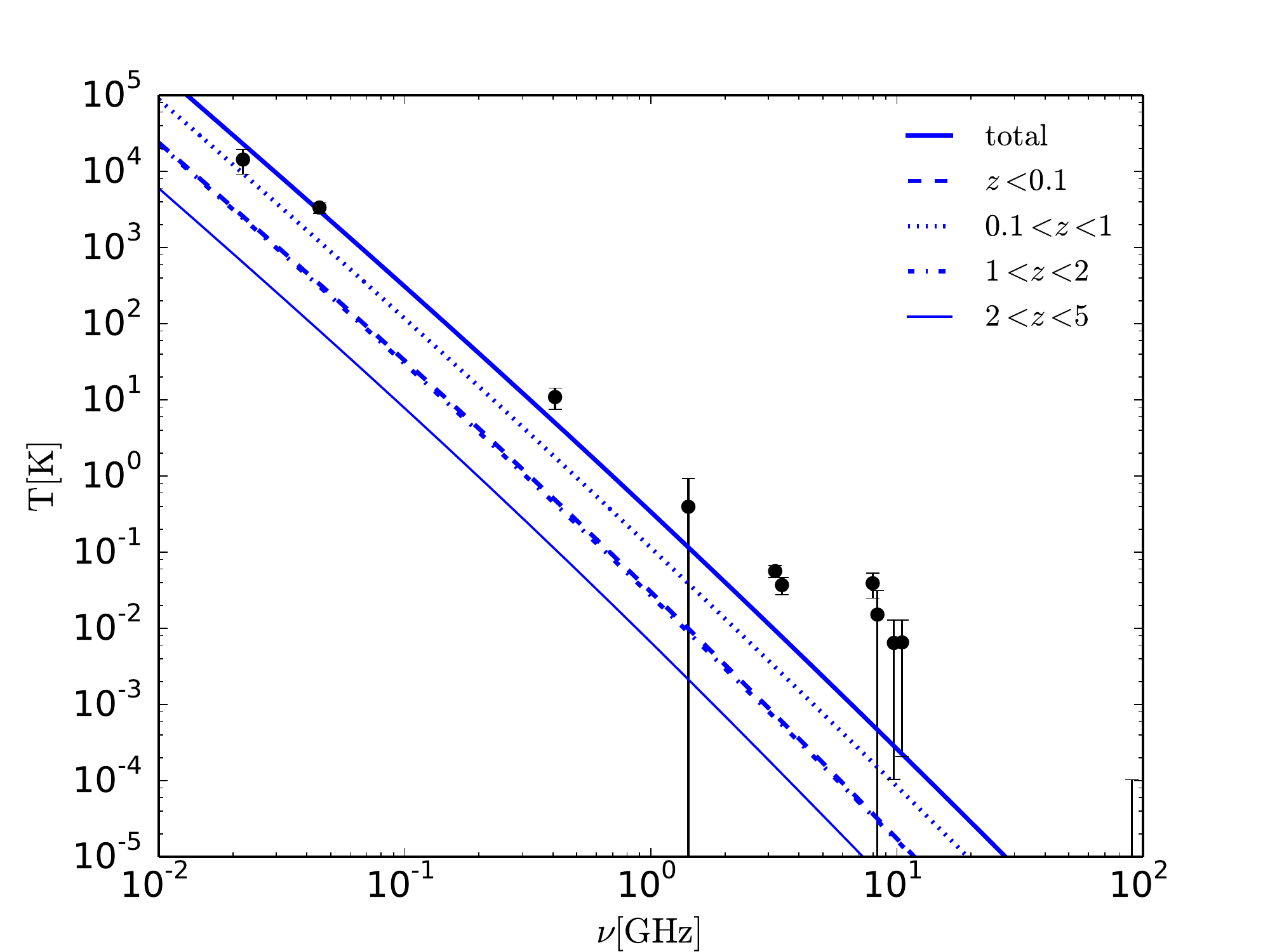,width=0.7\textwidth}   
\caption{\label{fig:yT_z} The sub-contributions to the total ARCADE-2 temperature stemming from halos at various distances. As in Figure~\ref{fig:yT_m},  $B_0 = 10\,\mu$G and  $\eta_v=0.86$ have been assumed for calculation. In this figure we  show the contributions from four redshift bins: $z<0.1$, $0.1<z<1$, $1<z<2$, and $2<z<5$.  }
\end{figure}

In Figure~\ref{fig:yT_m} we decompose the cumulative radio emission into sub-contributions from host halos in three mass groups: $M<10^{12}\,M_\odot$, $10^{12}\,M_\odot<M<10^{14}\,M_\odot$, and $M>10^{14}\,M_\odot$.  We find that the most massive group contributes dominantly with $\sim 60\%$,   the $10^{12}\,M_\odot<M<10^{14}\,M_\odot$ group contributes $40\%$,
and the least massive group contributes negligibly to  the observed radio excess. 
{One caveat is that the fractional contribution  from clusters in different mass groups   relies on the  scaling of magnetic field with the cluster mass  (as defined in Equation~\ref{eqn:B_cluster}) through the 
the $B$-dependent   $E_{\rm max}$ (Equation~\ref{eqn:Emax} and \ref{eqn:SA}). The partition of contributions of subgroups in Figure~\ref{fig:yT_m} could be altered   if  very different  $E_{\rm max}$ and $B(M)$ relations would be invoked.   }

Additionally, in Figure~\ref{fig:yT_z} we decompose the synchrotron signals into sub-contributions from host halos located at four different redshift bins:  $z<0.1$, $0.1<z<1$, $1<z<2$, and $2<z<5$.  It is found that the merger events at $0.1<z<1$ contribute  most significantly, as a result of the fact that massive clusters are most abundant in the nearby universe. 

Intriguingly, we note that the anisotropy produced by cluster mergers is also expected to be quite low. While we withhold on a quantitative calculation of the anisotropy, we note that a $10^{14}\,M_\odot$ cluster at $z=1$ has an angular size of $6'$, which corresponds to $\ell\sim 3600$. Therefore the dominate emission sources in our model (M$_L$~$>$~10$^{14}$~M$_\odot$ and z~$<$~1) have angular sizes significantly greater than the $2'$ constraint imposed by \cite{Vernstrom:2014uda}, and are beyond the range of the angular scales that could be probed by the radio measurements quoted in \cite{holder_anisotropy_of_arcade}.

\section{Discussion and Conclusions}
\label{sec:conclusions}

In this \emph{paper} we have shown that cluster mergers are capable of powering the bright isotropic synchrotron emission necessary to explain the ARCADE-2 excess. Specifically, these cluster mergers provide a number of advantages over current explanations from both astrophysical and dark matter models. The large power of the ARCADE-2 excess is easily provided by cluster mergers, and the spectrum of the ARCADE-2 excess is well-fit for steady-state electron spectra following ${dN}/{dE}$~$\propto$~$E^{-2.5}$ with a cut off at E$_{\rm max}~\sim$~50~GeV and a cluster magnetic field strength near the virial radius of $\sim$1~$\mu$G. It is enticing to note that the observed radio data is so easily fit by these parameter choices, which are among the more theoretically well-motivated and observationally constrained cluster parameters~\citep{2012SSRv..166..187B, 2012A&ARv..20...54F}. The turbulence and subsequent re-acceleration of non-thermal electrons is intrinsically a non-thermal process, which produces negligible IR and X-Ray emission, thus remaining consistent with high-energy constraints. Finally, the synchrotron emission is dominated by the largest clusters, significantly decreasing the predicted anisotropy from this model at high-multipoles.

There are two caveats to the analysis shown in this work. First, while our model predicts a maximum electron energy of $\sim$50~GeV, this stems from a simple, time-independent approximation that does not carefully take into account between Alfv{\'e}n reacceleration and energy-loss process such as synchrotron emission. A full magnetohydrodynamic model of cluster mergers is necessary to carefully model the electron energy spectrum in both minor and major mergers, an effort which lies beyond the scope of the current work. In Appendix A, we discuss this caveat in more detail, and show that reasonable variations in the maximum electron energy moderately impact our fit to the high-frequency excess, but do not effect fits below $\sim$1 GHz. Additionally, we have assumed that the power of Alfv\'en waves could be efficiently converted into the re-acceleration of electrons.  This assumption may not be guaranteed.  For instance, when the energy density of relativistic protons is larger than that of electrons, an efficient electron acceleration would unavoidably accelerate protons, leading to an  increase of the damping rate of the Alfv\'en waves and thus a decrease in   the efficiency of electron acceleration. However, as cosmic rays at GeV-TeV energies are expected to be confined within the cluster volume for cosmological times, re-energization of secondary electrons and positrons is possible due to the inelastic interactions between cosmic ray protons and the intracluster medium \citep{Brunetti05}.
 
Furthermore, in our model we  set the turbulence scale to be 2 times the stripping radius, which could underestimate $l_0$ especially in cases of minor mergers. For example, observations  show evidence of minor merging activity in the massive cluster Abell 2142 ($M_{200}=1.3\times10^{15}\,M_\odot$)  with a subhalo of mass $\sim10^{13}\,M_\odot$ \citep{2011ApJ...741..122O,2014A&A...570A.119E}. This merging activity is suggested to be inducing turbulence and creating an extended radio halo with a linear scale of 2 Mpc  \citep{2013ApJ...779..189F}, but in our model such a merger only generates turbulence for less than 200 kpc. The possibility of larger turbulent volume leaves  room for a  scenario with less efficient electron acceleration. On the contrary, if the actual turbulence volume is not as large as modelled in this work, or equivalently, if  on average the turbulence channels less than $20\%$ of thermal energy in major mergers, and $0.5\%$ in minor mergers, our model would have over estimated the synchrotron emission from cluster mergers. 


Finally, we note that our model provides a highly testable observational signature. Measurements of the radio anisotropy at larger angular scales would be capable of detecting a significant anisotropy stemming from cluster contributions to the ARCADE-2 excess.
Forthcoming observations with the Low Frequency Array (LOFAR) \citep{2013A&A...556A...2V}, the Murchison Widefield Array (MWA) \citep{2013PASA...30....7T}, and  the planned Square Kilometre Array (SKA) \citep{2009IEEEP..97.1482D} should be sufficient to detect such anisotropy signals \citep{2004ApJ...617..281K, 2012SSRv..166..215B}.  Moreover, the cross-correlation of this anisotropy measurement with existing catalogs of massive clusters  (e.g. SPT catalog \citep{Bleem:2014iim}, ACT survey \citep{2011ApJ...737...61M, 2013JCAP...07..008H}, Planck survey \citep{2014A&A...571A..29P}, ROSAT all-sky survey \citep{2011A&A...534A.109P}), would provide smoking gun evidence for cluster contributions to the isotropic radio background. Due to the expected presence of high degrees of turbulence in cluster shocks, the lack of significant radio emission in cluster mergers would also be of great interest, informing our understanding of re-acceleration in these energetic environments, and motivating the search for more exotic explanations of the ARCADE-2 data.

\appendix
\label{app:appendix}
\section{Impact of more realistic electron spectra}
\begin{figure}
\centering
\epsfig{file=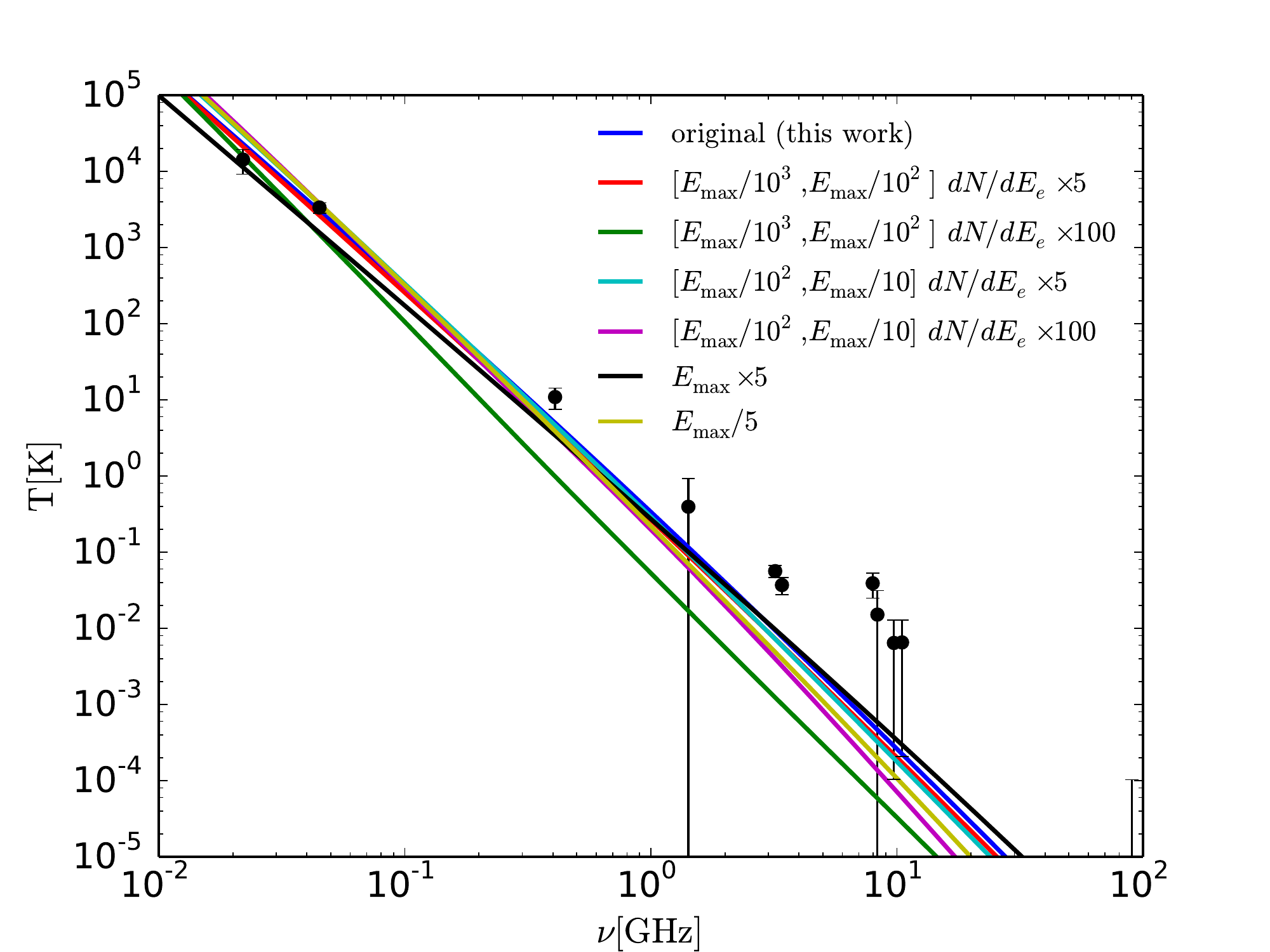,width=0.7\textwidth}   
\caption{\label{fig:MHD_compare} Comparison of  radio emission by different electron spectra. In addition to the power-law spectrum with $dN/dE_e\propto E^{-2.5}$ in  our default case (see equation~\ref{eqn:dNdEe}), we assume the presence of a bump with an amplitude N times higher  for the following energy ranges: 1)  $N=5$ at $[E_{\rm max}/10^3, E_{\rm max}/10^2]$,  2)  $N=100$ at $[E_{\rm max}/10^3, E_{\rm max}/10^2]$,  3)  $N=5$ at $[E_{\rm max}/10^2, E_{\rm max}/10]$  and  4)  $N=100$ at $[E_{\rm max}/10^2, E_{\rm max}/10]$.  We also show the impact to the fit  if  the maximum electron energy is 5 times greater or smaller than the $E_{\rm max}$ defined in equation~\ref{eqn:Emax}.  The modified spectra are normalized by the Alfv\'en injection power through equation~\ref{eq:P_SI}. 
In all cases the magnetic field  is assumed to   have a central value $B_0 = 10\,\mu$G. The turbulent velocity is assumed to be a fraction $\eta_v=0.86$  of the initial relative velocity of the merging clusters.
}
\end{figure}

Throughout this work we have adopted an analytical approach to calculate  the Alfv\'en re-acceleration following  \cite{1984ApJ...277..820E} and \cite{Fujita03}. In particular, we have  assumed that injection electrons follow a  simple power-law of $dN/dE_e\propto E^{-2.5}$ (see equation~\ref{eqn:dNdEe}). However,  numerical simulations suggested a more comprehensive electron spectrum with  a characteristic bump slightly before the cut-off energy (see for example, Figure~3 of \cite{Brunetti05}), as a result of  the interplay of the turbulent re-acceleration of electrons and protons, as well as the inelastic interaction between cosmic ray protons and the ICM. To evaluate the impact of more realistic electron spectra, below we test our model by adding artificial bumps to the injection spectrum to mimic the numerical results. Specifically, we assume that  a bump with $dN/dE_e$  N times higher than the original power-law flux is placed  between $E_{\rm max}/c_{\rm l}$ and $ E_{\rm max}/c_{\rm r}$. The modified spectrum is then normalized by the Alfv\'en injection power through equation~\ref{eq:P_SI}. 
Figure~\ref{fig:MHD_compare} present the   synchrotron emission from the following six tests:  1) N = 5, $c_l=10^3$, $c_r=10^2$ 2) N = 100, $c_l=10^3$, $c_r=10^2$ 3) N = 5, $c_l=10^2$, $c_r=10$ 4) N = 100, $c_l=10^2$, $c_r=10$.  
 The spectrum shows a significant derivation from the initial result in case 2), where most electrons have low energies and thus produce a much softer spectrum compared to observation. However for all the other cases, after integrated over the entire halo population, the bumps are smoothed out and produce very small changes to our results.  The above tests show that our conclusions are robust under different inputs of electron spectra. 

 In addition, to test the dependence of our results on the maximum energy of the re-accelerated electrons, we artificially injected spectra with  maximum energies 5 times higher or lower than the $E_{\rm max}$ calculated by equation~\ref{eqn:Emax}. As shown by the black and yellow lines in Figure~\ref{fig:MHD_compare}, $E_{\rm max}$ could slightly modify the overall synchrotron spectrum, but the effect is rather minor.  
  On the other hand,  if $E_{\rm max}$ would have been significantly less than  the value from equation~\ref{eqn:Emax}, the resulting spectrum would be  even softer, and it may become difficult to explain the high-frequency tail of the ARCADE-2 emission. A  more advanced  MHD modelling would be important to test the impact from a time-dependent $E_{\rm max}$.

\section{Magnetic energy compared to the thermal energy  }

\begin{figure}
\centering
\epsfig{file=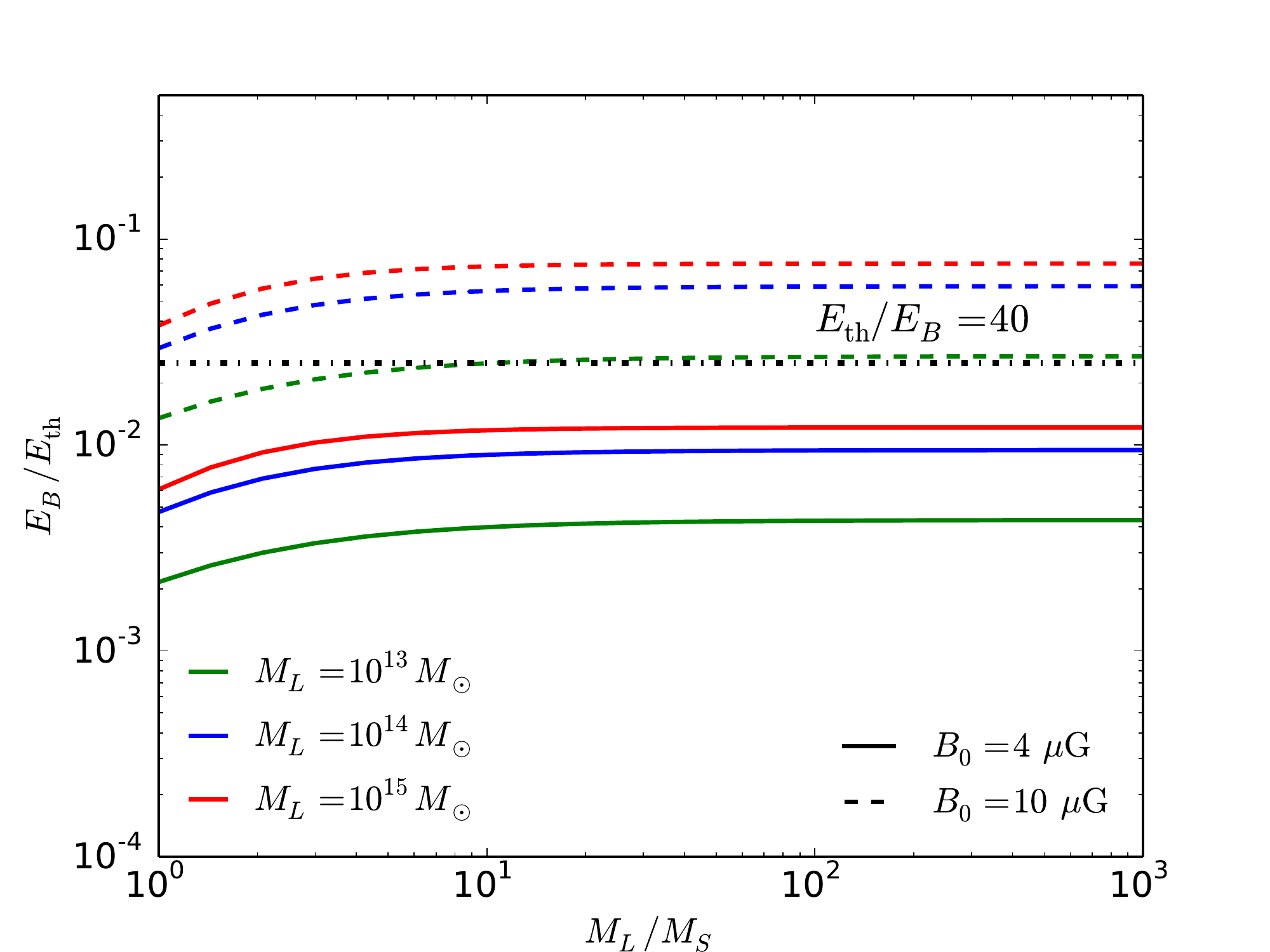,width=0.7\textwidth}   
\caption{\label{fig:AveEBOverEth} The  ratio of the magnetic energy $E_B$   to the total thermal energy $E_{\rm th}$ in the turbulent volume as a function of the mass ratios, for a central magnetic field strength $B_0=4\,\mu$G (solid lines), and $B_0=10\,\mu$G (dashed lines). For reference, the black dash-dotted line indicates the level  $E_{\rm th}/E_B\approx 40$ as suggested  in  \cite{Miniati:2015gga}. }
\end{figure}

As a consistency check, in Figure~\ref{fig:AveEBOverEth} we estimate the ratio of the magnetic energy, $E_B$    to the total thermal energy $E_{\rm th}$ of the merging clusters in the turbulent volume. Specifically, the ratio is calculated by 
\bey
\frac{E_B}{E_{\rm th}} (M_L, M_S, z) =\left(  \int \frac{B^2}{8\pi}\,dV_t \right) /\left(\frac{(M_L+M_S)\, k_B\,T_L}{\mu \,m_p}\, \frac{V_t}{V_L}\right)
\eey
The ratios reported in figure~\ref{fig:AveEBOverEth}  range from 2\%  to 8\% for $B_0=10\,\mu$G, and from 0.2\% to 1.1\% for $B_0=4\,\mu$G. These estimated values   are comparable with  the level ($\sim 2.5\%$)  suggested in \cite{Miniati:2015gga}.

\acknowledgments 
The authors acknowledge Dan Hooper, Alex Lazarian, and Angela Olinto for helpful discussions. KF acknowledges financial support from the NSF grant PHY-1068696 and  the NASA grant 11-APRA-0066 at the University of Chicago, and the grant NSF PHY-1125897 at the Kavli Institute for Cosmological Physics. KF acknowledges the support of a Joint Space-Science Institute prize postdoctoral fellowship.  TL is supported by the National Aeronautics and Space Administration through Einstein Postdoctoral Fellowship Award Number PF3-140110.  This work made use of computing resources and support provided by the Research Computing Center at the University of Chicago.

\bibliography{FL16} 

\providecommand{\href}[2]{#2}\begingroup\raggedright\begin{thebibliography}{10}

\bibitem{haslam_408Mhz_excess}
C.~G.~T. {Haslam}, U.~{Klein}, C.~J. {Salter}, H.~{Stoffel}, W.~E. {Wilson},
  M.~N. {Cleary} et~al., \emph{{A 408 MHz all-sky continuum survey. I -
  Observations at southern declinations and for the North Polar region}},
  {\emph{AAP} {\bf 100} (July, 1981) 209--219}.

\bibitem{roger_22MHz_excess}
R.~S. {Roger}, C.~H. {Costain}, T.~L. {Landecker} and C.~M. {Swerdlyk},
  \emph{{The radio emission from the Galaxy at 22 MHz}},
  \href{http://dx.doi.org/10.1051/aas:1999239}{\emph{AAPS} {\bf 137} (May,
  1999) 7--19}, [\href{http://arxiv.org/abs/arXiv:astro-ph/9902213}{{\tt
  arXiv:astro-ph/9902213}}].

\bibitem{reich_1.4GHz_excess}
P.~{Reich} and W.~{Reich}, \emph{{A radio continuum survey of the northern sky
  at 1420 MHz. II}}, {\emph{AAPS} {\bf 63} (Feb., 1986) 205--288}.

\bibitem{guzman_44Mhz_excess}
A.~E. {Guzm{\'a}n}, J.~{May}, H.~{Alvarez} and K.~{Maeda}, \emph{{All-sky
  Galactic radiation at 45 MHz and spectral index between 45 and 408 MHz}},
  \href{http://dx.doi.org/10.1051/0004-6361/200913628}{\emph{AAP} {\bf 525}
  (Jan., 2011) A138}, [\href{http://arxiv.org/abs/1011.4298}{{\tt 1011.4298}}].

\bibitem{arcade_interpretation}
M.~{Seiffert}, D.~J. {Fixsen}, A.~{Kogut}, S.~M. {Levin}, M.~{Limon}, P.~M.
  {Lubin} et~al., \emph{{Interpretation of the Extragalactic Radio
  Background}}, {\emph{ArXiv e-prints} (Jan., 2009) },
  [\href{http://arxiv.org/abs/0901.0559}{{\tt 0901.0559}}].

\bibitem{arcade_measurement}
D.~J. {Fixsen}, A.~{Kogut}, S.~{Levin}, M.~{Limon}, P.~{Lubin}, P.~{Mirel}
  et~al., \emph{{ARCADE 2 Measurement of the Extra-Galactic Sky Temperature at
  3-90 GHz}}, {\emph{ArXiv e-prints} (Jan., 2009) },
  [\href{http://arxiv.org/abs/0901.0555}{{\tt 0901.0555}}].

\bibitem{2014JCAP...04..008F}
N.~{Fornengo}, R.~A. {Lineros}, M.~{Regis} and M.~{Taoso}, \emph{{The isotropic
  radio background revisited}},
  \href{http://dx.doi.org/10.1088/1475-7516/2014/04/008}{\emph{JCAP} {\bf 4}
  (Apr., 2014) 8}, [\href{http://arxiv.org/abs/1402.2218}{{\tt 1402.2218}}].

\bibitem{kogut_excess_not_galactic}
A.~{Kogut}, D.~J. {Fixsen}, S.~M. {Levin}, M.~{Limon}, P.~M. {Lubin},
  P.~{Mirel} et~al., \emph{{ARCADE 2 Observations of Galactic Radio Emission}},
  \href{http://dx.doi.org/10.1088/0004-637X/734/1/4}{\emph{ApJ} {\bf 734}
  (June, 2011) 4}, [\href{http://arxiv.org/abs/0901.0562}{{\tt 0901.0562}}].

\bibitem{singal_not_extragalactic_baryonic_signals}
J.~{Singal}, {\L}.~{Stawarz}, A.~{Lawrence} and V.~{Petrosian}, \emph{{Sources
  of the radio background considered}},
  \href{http://dx.doi.org/10.1111/j.1365-2966.2010.17382.x}{\emph{MNRAS} {\bf
  409} (Dec., 2010) 1172--1182}, [\href{http://arxiv.org/abs/0909.1997}{{\tt
  0909.1997}}].

\bibitem{gervasi_radio_luminosity_from_extrapolation_of_known_sources}
M.~{Gervasi}, A.~{Tartari}, M.~{Zannoni}, G.~{Boella} and G.~{Sironi},
  \emph{{The Contribution of the Unresolved Extragalactic Radio Sources to the
  Brightness Temperature of the Sky}},
  \href{http://dx.doi.org/10.1086/588628}{\emph{ApJ} {\bf 682} (July, 2008)
  223--230}, [\href{http://arxiv.org/abs/0803.4138}{{\tt 0803.4138}}].

\bibitem{fornengo_arcade_excess_is_dm}
N.~{Fornengo}, R.~{Lineros}, M.~{Regis} and M.~{Taoso}, \emph{{Possibility of a
  Dark Matter Interpretation for the Excess in Isotropic Radio Emission
  Reported by ARCADE}},
  \href{http://dx.doi.org/10.1103/PhysRevLett.107.271302}{\emph{Physical Review
  Letters} {\bf 107} (Dec., 2011) A261302},
  [\href{http://arxiv.org/abs/1108.0569}{{\tt 1108.0569}}].

\bibitem{fornengo_arcade_excess_is_dm2}
N.~{Fornengo}, R.~{Lineros}, M.~{Regis} and M.~{Taoso}, \emph{{Cosmological
  radio emission induced by WIMP Dark Matter}},
  \href{http://dx.doi.org/10.1088/1475-7516/2012/03/033}{\emph{JCAP} {\bf 3}
  (Mar., 2012) 33}, [\href{http://arxiv.org/abs/1112.4517}{{\tt 1112.4517}}].

\bibitem{hooper_arcade_excess}
D.~{Hooper}, A.~V. {Belikov}, T.~E. {Jeltema}, T.~{Linden}, S.~{Profumo} and
  T.~R. {Slatyer}, \emph{{The isotropic radio background and annihilating dark
  matter}}, \href{http://dx.doi.org/10.1103/PhysRevD.86.103003}{\emph{PRD} {\bf
  86} (Nov., 2012) 103003}, [\href{http://arxiv.org/abs/1203.3547}{{\tt
  1203.3547}}].

\bibitem{FL14}
K.~Fang and T.~Linden, \emph{{Anisotropy of the extragalactic radio background
  from dark matter annihilation}},
  \href{http://dx.doi.org/10.1103/PhysRevD.91.083501}{\emph{Phys.Rev.} {\bf
  D91} (2015) 083501}, [\href{http://arxiv.org/abs/1412.7545}{{\tt
  1412.7545}}].

\bibitem{holder_anisotropy_of_arcade}
G.~{Holder}, \emph{{The unusual smoothness of the extragalactic unresolved
  radio background}}, {\emph{ArXiv e-prints} (July, 2012) },
  [\href{http://arxiv.org/abs/1207.0856}{{\tt 1207.0856}}].

\bibitem{Vernstrom:2014uda}
T.~Vernstrom, R.~P. Norris, D.~Scott and J.~Wall, \emph{{The Deep Diffuse
  Extragalactic Radio Sky at 1.75 GHz}},
  \href{http://arxiv.org/abs/1408.4160}{{\tt 1408.4160}}.

\bibitem{2002ARA&A..40..319C}
C.~L. {Carilli} and G.~B. {Taylor}, \emph{{Cluster Magnetic Fields}},
  \href{http://dx.doi.org/10.1146/annurev.astro.40.060401.093852}{\emph{Annual
  Review of Astronomy and Astrophysics} {\bf 40} (2002) 319--348},
  [\href{http://arxiv.org/abs/astro-ph/0110655}{{\tt astro-ph/0110655}}].

\bibitem{2008MNRAS.391.1685S}
V.~{Springel}, J.~{Wang}, M.~{Vogelsberger}, A.~{Ludlow}, A.~{Jenkins},
  A.~{Helmi} et~al., \emph{{The Aquarius Project: the subhaloes of galactic
  haloes}},
  \href{http://dx.doi.org/10.1111/j.1365-2966.2008.14066.x}{\emph{MNRAS} {\bf
  391} (Dec., 2008) 1685--1711}, [\href{http://arxiv.org/abs/0809.0898}{{\tt
  0809.0898}}].

\bibitem{2012MNRAS.419.1721G}
L.~{Gao}, C.~S. {Frenk}, A.~{Jenkins}, V.~{Springel} and S.~D.~M. {White},
  \emph{{Where will supersymmetric dark matter first be seen?}},
  \href{http://dx.doi.org/10.1111/j.1365-2966.2011.19836.x}{\emph{MNRAS} {\bf
  419} (Jan., 2012) 1721--1726}, [\href{http://arxiv.org/abs/1107.1916}{{\tt
  1107.1916}}].

\bibitem{2012MNRAS.425.2169G}
L.~{Gao}, J.~F. {Navarro}, C.~S. {Frenk}, A.~{Jenkins}, V.~{Springel} and
  S.~D.~M. {White}, \emph{{The Phoenix Project: the dark side of rich Galaxy
  clusters}},
  \href{http://dx.doi.org/10.1111/j.1365-2966.2012.21564.x}{\emph{MNRAS} {\bf
  425} (Sept., 2012) 2169--2186}, [\href{http://arxiv.org/abs/1201.1940}{{\tt
  1201.1940}}].

\bibitem{2014MNRAS.442.2271S}
M.~A. {S{\'a}nchez-Conde} and F.~{Prada}, \emph{{The flattening of the
  concentration-mass relation towards low halo masses and its implications for
  the annihilation signal boost}},
  \href{http://dx.doi.org/10.1093/mnras/stu1014}{\emph{MNRAS} {\bf 442} (Aug.,
  2014) 2271--2277}, [\href{http://arxiv.org/abs/1312.1729}{{\tt 1312.1729}}].

\bibitem{2014IJMPD..2330007B}
G.~{Brunetti} and T.~W. {Jones}, \emph{{Cosmic Rays in Galaxy Clusters and
  Their Nonthermal Emission}},
  \href{http://dx.doi.org/10.1142/S0218271814300079}{\emph{International
  Journal of Modern Physics D} {\bf 23} (Mar., 2014) 1430007--98},
  [\href{http://arxiv.org/abs/1401.7519}{{\tt 1401.7519}}].

\bibitem{1984ApJ...277..820E}
J.~A. {Eilek} and R.~N. {Henriksen}, \emph{{The electron energy spectrum
  produced in radio sources by turbulent, resonant acceleration}},
  \href{http://dx.doi.org/10.1086/161752}{\emph{ApJ} {\bf 277} (Feb., 1984)
  820--831}.

\bibitem{Fujita03}
Y.~{Fujita}, M.~{Takizawa} and C.~L. {Sarazin}, \emph{{Nonthermal Emissions
  from Particles Accelerated by Turbulence in Clusters of Galaxies}},
  \href{http://dx.doi.org/10.1086/345599}{\emph{ApJ} {\bf 584} (Feb., 2003)
  190--202}, [\href{http://arxiv.org/abs/astro-ph/0210320}{{\tt
  astro-ph/0210320}}].

\bibitem{1965PhFl....8.1385K}
R.~H. {Kraichnan}, \emph{{Inertial-Range Spectrum of Hydromagnetic
  Turbulence}}, \href{http://dx.doi.org/10.1063/1.1761412}{\emph{Physics of
  Fluids} {\bf 8} (July, 1965) 1385--1387}.

\bibitem{1952RSPSA.211..564L}
M.~J. {Lighthill}, \emph{{On Sound Generated Aerodynamically. I. General
  Theory}}, \href{http://dx.doi.org/10.1098/rspa.1952.0060}{\emph{Royal Society
  of London Proceedings Series A} {\bf 211} (Mar., 1952) 564--587}.

\bibitem{1955ApJ...121..461K}
R.~M. {Kulsrud}, \emph{{Effect of Magnetic Fields on Generation of Noise by
  Isotropic Turbulence.}}, \href{http://dx.doi.org/10.1086/146008}{\emph{ApJ}
  {\bf 121} (Mar., 1955) 461}.

\bibitem{1962pfig.book.....S}
L.~{Spitzer}, \emph{{Physics of Fully Ionized Gases}}.
\newblock 1962.

\bibitem{1942Natur.150..405A}
H.~{Alfv{\'e}n}, \emph{{Existence of Electromagnetic-Hydrodynamic Waves}},
  \href{http://dx.doi.org/10.1038/150405d0}{\emph{Nature} {\bf 150} (Oct.,
  1942) 405--406}.

\bibitem{Brunetti04}
G.~{Brunetti}, P.~{Blasi}, R.~{Cassano} and S.~{Gabici}, \emph{{Alfv{\'e}nic
  reacceleration of relativistic particles in galaxy clusters: MHD waves,
  leptons and hadrons}},
  \href{http://dx.doi.org/10.1111/j.1365-2966.2004.07727.x}{\emph{MNRAS} {\bf
  350} (June, 2004) 1174--1194},
  [\href{http://arxiv.org/abs/astro-ph/0312482}{{\tt astro-ph/0312482}}].

\bibitem{1979ApJ...230..373E}
J.~A. {Eilek}, \emph{{Particle reacceleration in radio galaxies}},
  \href{http://dx.doi.org/10.1086/157093}{\emph{ApJ} {\bf 230} (June, 1979)
  373--385}.

\bibitem{1992ApJ...398..350H}
R.~J. {Hamilton} and V.~{Petrosian}, \emph{{Stochastic acceleration of
  electrons. I - Effects of collisions in solar flares}},
  \href{http://dx.doi.org/10.1086/171860}{\emph{ApJ} {\bf 398} (Oct., 1992)
  350--358}.

\bibitem{2000ApJ...535..586T}
M.~{Takizawa} and T.~{Naito}, \emph{{Nonthermal Emission from Relativistic
  Electrons in Clusters of Galaxies: A Merger Shock Acceleration Model}},
  \href{http://dx.doi.org/10.1086/308894}{\emph{ApJ} {\bf 535} (June, 2000)
  586--592}, [\href{http://arxiv.org/abs/astro-ph/0001046}{{\tt
  astro-ph/0001046}}].

\bibitem{2009MNRAS.395.1333V}
F.~{Vazza}, G.~{Brunetti} and C.~{Gheller}, \emph{{Shock waves in Eulerian
  cosmological simulations: main properties and acceleration of cosmic rays}},
  \href{http://dx.doi.org/10.1111/j.1365-2966.2009.14691.x}{\emph{MNRAS} {\bf
  395} (May, 2009) 1333--1354}, [\href{http://arxiv.org/abs/0808.0609}{{\tt
  0808.0609}}].

\bibitem{1997ApJ...477..560E}
T.~A. {Ensslin}, P.~L. {Biermann}, P.~P. {Kronberg} and X.-P. {Wu},
  \emph{{Cosmic-Ray Protons and Magnetic Fields in Clusters of Galaxies and
  Their Cosmological Consequences}}, {\emph{ApJ} {\bf 477} (Mar., 1997)
  560--567}, [\href{http://arxiv.org/abs/astro-ph/9609190}{{\tt
  astro-ph/9609190}}].

\bibitem{1999APh....11...73V}
H.~J. {V{\"o}lk} and A.~M. {Atoyan}, \emph{{Clusters of galaxies: magnetic
  fields and nonthermal emission}},
  \href{http://dx.doi.org/10.1016/S0927-6505(99)00029-8}{\emph{Astroparticle
  Physics} {\bf 11} (June, 1999) 73--82},
  [\href{http://arxiv.org/abs/astro-ph/9812458}{{\tt astro-ph/9812458}}].

\bibitem{1968Ap&SS...2..171M}
D.~B. {Melrose}, \emph{{The Emission and Absorption of Waves by Charged
  Particles in Magnetized Plasmas}},
  \href{http://dx.doi.org/10.1007/BF00651567}{\emph{Astrophysics and Space
  Science} {\bf 2} (Oct., 1968) 171--235}.

\bibitem{Brunetti05}
G.~{Brunetti} and P.~{Blasi}, \emph{{Alfv{\'e}nic reacceleration of
  relativistic particles in galaxy clusters in the presence of secondary
  electrons and positrons}},
  \href{http://dx.doi.org/10.1111/j.1365-2966.2005.09511.x}{\emph{MNRAS} {\bf
  363} (Nov., 2005) 1173--1187},
  [\href{http://arxiv.org/abs/astro-ph/0508100}{{\tt astro-ph/0508100}}].

\bibitem{1995ApJ...438..763G}
P.~{Goldreich} and S.~{Sridhar}, \emph{{Toward a theory of interstellar
  turbulence. 2: Strong alfvenic turbulence}},
  \href{http://dx.doi.org/10.1086/175121}{\emph{ApJ} {\bf 438} (Jan., 1995)
  763--775}.

\bibitem{2003ApJ...595..812C}
J.~{Cho}, A.~{Lazarian} and E.~T. {Vishniac}, \emph{{Ordinary and
  Viscosity-damped Magnetohydrodynamic Turbulence}},
  \href{http://dx.doi.org/10.1086/377515}{\emph{ApJ} {\bf 595} (Oct., 2003)
  812--823}, [\href{http://arxiv.org/abs/astro-ph/0305212}{{\tt
  astro-ph/0305212}}].

\bibitem{2006ApJ...645L..25L}
A.~{Lazarian}, \emph{{Enhancement and Suppression of Heat Transfer by MHD
  Turbulence}}, \href{http://dx.doi.org/10.1086/505796}{\emph{ApJL} {\bf 645}
  (July, 2006) L25--L28}, [\href{http://arxiv.org/abs/astro-ph/0608045}{{\tt
  astro-ph/0608045}}].

\bibitem{2000PhRvL..85.4656C}
B.~D.~G. {Chandran}, \emph{{Scattering of Energetic Particles by Anisotropic
  Magnetohydrodynamic Turbulence with a Goldreich-Sridhar Power Spectrum}},
  \href{http://dx.doi.org/10.1103/PhysRevLett.85.4656}{\emph{Physical Review
  Letters} {\bf 85} (Nov., 2000) 4656--4659},
  [\href{http://arxiv.org/abs/astro-ph/0008498}{{\tt astro-ph/0008498}}].

\bibitem{2007MNRAS.378..245B}
G.~{Brunetti} and A.~{Lazarian}, \emph{{Compressible turbulence in galaxy
  clusters: physics and stochastic particle re-acceleration}},
  \href{http://dx.doi.org/10.1111/j.1365-2966.2007.11771.x}{\emph{MNRAS} {\bf
  378} (June, 2007) 245--275},
  [\href{http://arxiv.org/abs/astro-ph/0703591}{{\tt astro-ph/0703591}}].

\bibitem{2015ApJ...800...60M}
F.~{Miniati}, \emph{{The Matryoshka Run. II. Time-dependent Turbulence
  Statistics, Stochastic Particle Acceleration, and Microphysics Impact in a
  Massive Galaxy Cluster}},
  \href{http://dx.doi.org/10.1088/0004-637X/800/1/60}{\emph{ApJ} {\bf 800}
  (Feb., 2015) 60}, [\href{http://arxiv.org/abs/1409.3576}{{\tt 1409.3576}}].

\bibitem{1976A&A....49..137C}
A.~{Cavaliere} and R.~{Fusco-Femiano}, \emph{{X-rays from hot plasma in
  clusters of galaxies}}, {\emph{A\&A} {\bf 49} (May, 1976) 137--144}.

\bibitem{2003ApJ...584..702H}
W.~{Hu} and A.~V. {Kravtsov}, \emph{{Sample Variance Considerations for Cluster
  Surveys}}, \href{http://dx.doi.org/10.1086/345846}{\emph{ApJ} {\bf 584}
  (Feb., 2003) 702--715}, [\href{http://arxiv.org/abs/astro-ph/0203169}{{\tt
  astro-ph/0203169}}].

\bibitem{2013ApJ...778...14G}
A.~H. {Gonzalez}, S.~{Sivanandam}, A.~I. {Zabludoff} and D.~{Zaritsky},
  \emph{{Galaxy Cluster Baryon Fractions Revisited}},
  \href{http://dx.doi.org/10.1088/0004-637X/778/1/14}{\emph{ApJ} {\bf 778}
  (Nov., 2013) 14}, [\href{http://arxiv.org/abs/1309.3565}{{\tt 1309.3565}}].

\bibitem{Dolag:2001vy}
K.~Dolag, S.~Schindler, F.~Govoni and L.~Feretti, \emph{{Correlation of the
  magnetic field and the intra-cluster gas density in galaxy clusters}},
  \href{http://dx.doi.org/10.1051/0004-6361:20011219}{\emph{Astron. Astrophys.}
  {\bf 378} (2001) 777}, [\href{http://arxiv.org/abs/astro-ph/0108485}{{\tt
  astro-ph/0108485}}].

\bibitem{Donnert:2008sn}
J.~Donnert, K.~Dolag, H.~Lesch and E.~Muller, \emph{{Cluster Magnetic Fields
  from Galactic Outflows}},
  \href{http://dx.doi.org/10.1111/j.1365-2966.2008.14132.x}{\emph{Mon. Not.
  Roy. Astron. Soc.} {\bf 392} (2009) 1008--1021},
  [\href{http://arxiv.org/abs/0808.0919}{{\tt 0808.0919}}].

\bibitem{2010A&A...513A..30B}
A.~{Bonafede}, L.~{Feretti}, M.~{Murgia}, F.~{Govoni}, G.~{Giovannini},
  D.~{Dallacasa} et~al., \emph{{The Coma cluster magnetic field from Faraday
  rotation measures}},
  \href{http://dx.doi.org/10.1051/0004-6361/200913696}{\emph{A\&A} {\bf 513}
  (Apr., 2010) A30}, [\href{http://arxiv.org/abs/1002.0594}{{\tt 1002.0594}}].

\bibitem{Govoni:2004as}
F.~Govoni and L.~Feretti, \emph{{Magnetic field in clusters of galaxies}},
  \href{http://dx.doi.org/10.1142/S0218271804005080}{\emph{Int. J. Mod. Phys.}
  {\bf D13} (2004) 1549--1594},
  [\href{http://arxiv.org/abs/astro-ph/0410182}{{\tt astro-ph/0410182}}].

\bibitem{2005JCAP...01..009D}
K.~{Dolag}, D.~{Grasso}, V.~{Springel} and I.~{Tkachev}, \emph{{Constrained
  simulations of the magnetic field in the local Universe and the propagation
  of ultrahigh energy cosmic rays}},
  \href{http://dx.doi.org/10.1088/1475-7516/2005/01/009}{\emph{JCAP} {\bf 1}
  (Jan., 2005) 009}, [\href{http://arxiv.org/abs/astro-ph/0410419}{{\tt
  astro-ph/0410419}}].

\bibitem{2002ASSL..272.....F}
L.~{Feretti}, I.~M. {Gioia} and G.~{Giovannini}, eds., \emph{{Merging Processes
  in Galaxy Clusters}}, vol.~272 of \emph{Astrophysics and Space Science
  Library}, June, 2002.
\newblock 10.1007/0-306-48096-4.

\bibitem{2012SSRv..166..187B}
M.~{Br{\"u}ggen}, A.~{Bykov}, D.~{Ryu} and H.~{R{\"o}ttgering}, \emph{{Magnetic
  Fields, Relativistic Particles, and Shock Waves in Cluster Outskirts}},
  \href{http://dx.doi.org/10.1007/s11214-011-9785-9}{\emph{Space Science
  Reviews} {\bf 166} (May, 2012) 187--213},
  [\href{http://arxiv.org/abs/1107.5223}{{\tt 1107.5223}}].

\bibitem{2012A&ARv..20...54F}
L.~{Feretti}, G.~{Giovannini}, F.~{Govoni} and M.~{Murgia}, \emph{{Clusters of
  galaxies: observational properties of the diffuse radio emission}},
  \href{http://dx.doi.org/10.1007/s00159-012-0054-z}{\emph{The Astronomy and
  Astrophysics Review} {\bf 20} (May, 2012) 54},
  [\href{http://arxiv.org/abs/1205.1919}{{\tt 1205.1919}}].

\bibitem{2005MNRAS.357.1313C}
R.~{Cassano} and G.~{Brunetti}, \emph{{Cluster mergers and non-thermal
  phenomena: a statistical magneto-turbulent model}},
  \href{http://dx.doi.org/10.1111/j.1365-2966.2005.08747.x}{\emph{MNRAS} {\bf
  357} (Mar., 2005) 1313--1329},
  [\href{http://arxiv.org/abs/astro-ph/0412475}{{\tt astro-ph/0412475}}].

\bibitem{2002ASSL..272....1S}
C.~L. {Sarazin}, \emph{{The Physics of Cluster Mergers}},  in \emph{Merging
  Processes in Galaxy Clusters} (L.~{Feretti}, I.~M. {Gioia} and
  G.~{Giovannini}, eds.), vol.~272 of \emph{Astrophysics and Space Science
  Library}, pp.~1--38, June, 2002.
\newblock \href{http://arxiv.org/abs/astro-ph/0105418}{{\tt astro-ph/0105418}}.
\newblock \href{http://dx.doi.org/10.1007/0-306-48096-4_1}{DOI}.

\bibitem{2006MNRAS.366.1437S}
K.~{Subramanian}, A.~{Shukurov} and N.~E.~L. {Haugen}, \emph{{Evolving
  turbulence and magnetic fields in galaxy clusters}},
  \href{http://dx.doi.org/10.1111/j.1365-2966.2006.09918.x}{\emph{MNRAS} {\bf
  366} (Mar., 2006) 1437--1454},
  [\href{http://arxiv.org/abs/astro-ph/0505144}{{\tt astro-ph/0505144}}].

\bibitem{2011A&A...529A..17V}
F.~{Vazza}, G.~{Brunetti}, C.~{Gheller}, R.~{Brunino} and M.~{Br{\"u}ggen},
  \emph{{Massive and refined. II. The statistical properties of turbulent
  motions in massive galaxy clusters with high spatial resolution}},
  \href{http://dx.doi.org/10.1051/0004-6361/201016015}{\emph{A\&A} {\bf 529}
  (May, 2011) A17}, [\href{http://arxiv.org/abs/1010.5950}{{\tt 1010.5950}}].

\bibitem{2012A&A...544A.103V}
F.~{Vazza}, E.~{Roediger} and M.~{Br{\"u}ggen}, \emph{{Turbulence in the ICM
  from mergers, cool-core sloshing, and jets: results from a new multi-scale
  filtering approach}},
  \href{http://dx.doi.org/10.1051/0004-6361/201118688}{\emph{A\&A} {\bf 544}
  (Aug., 2012) A103}, [\href{http://arxiv.org/abs/1202.5882}{{\tt 1202.5882}}].

\bibitem{2015ASSL..407..557B}
G.~{Brunetti} and T.~W. {Jones}, \emph{{Cosmic Rays in Galaxy Clusters and
  Their Interaction with Magnetic Fields}},  in \emph{Astrophysics and Space
  Science Library} (A.~{Lazarian}, E.~M. {de Gouveia Dal Pino} and
  C.~{Melioli}, eds.), vol.~407 of \emph{Astrophysics and Space Science
  Library}, p.~557, 2015.
\newblock \href{http://dx.doi.org/10.1007/978-3-662-44625-6_20}{DOI}.

\bibitem{2004A&A...426..387S}
P.~{Schuecker}, A.~{Finoguenov}, F.~{Miniati}, H.~{B{\"o}hringer} and U.~G.
  {Briel}, \emph{{Probing turbulence in the Coma galaxy cluster}},
  \href{http://dx.doi.org/10.1051/0004-6361:20041039}{\emph{A\&A} {\bf 426}
  (Nov., 2004) 387--397}, [\href{http://arxiv.org/abs/astro-ph/0404132}{{\tt
  astro-ph/0404132}}].

\bibitem{2008ApJ...688..709T}
J.~{Tinker}, A.~V. {Kravtsov}, A.~{Klypin}, K.~{Abazajian}, M.~{Warren},
  G.~{Yepes} et~al., \emph{{Toward a Halo Mass Function for Precision
  Cosmology: The Limits of Universality}},
  \href{http://dx.doi.org/10.1086/591439}{\emph{ApJ} {\bf 688} (Dec., 2008)
  709--728}, [\href{http://arxiv.org/abs/0803.2706}{{\tt 0803.2706}}].

\bibitem{1999ApJ...511....5E}
D.~J. {Eisenstein} and W.~{Hu}, \emph{{Power Spectra for Cold Dark Matter and
  Its Variants}}, \href{http://dx.doi.org/10.1086/306640}{\emph{ApJ} {\bf 511}
  (Jan., 1999) 5--15}, [\href{http://arxiv.org/abs/astro-ph/9710252}{{\tt
  astro-ph/9710252}}].

\bibitem{sheth_tormen_mass_function}
R.~K. {Sheth} and G.~{Tormen}, \emph{{Large-scale bias and the peak background
  split}},
  \href{http://dx.doi.org/10.1046/j.1365-8711.1999.02692.x}{\emph{MNRAS} {\bf
  308} (Sept., 1999) 119--126},
  [\href{http://arxiv.org/abs/arXiv:astro-ph/9901122}{{\tt
  arXiv:astro-ph/9901122}}].

\bibitem{2011ApJ...741..122O}
M.~S. {Owers}, P.~E.~J. {Nulsen} and W.~J. {Couch}, \emph{{Minor Merger-induced
  Cold Fronts in Abell 2142 and RXJ1720.1+2638}},
  \href{http://dx.doi.org/10.1088/0004-637X/741/2/122}{\emph{ApJ} {\bf 741}
  (Nov., 2011) 122}, [\href{http://arxiv.org/abs/1109.5692}{{\tt 1109.5692}}].

\bibitem{2014A&A...570A.119E}
D.~{Eckert}, S.~{Molendi}, M.~{Owers}, M.~{Gaspari}, T.~{Venturi}, L.~{Rudnick}
  et~al., \emph{{The stripping of a galaxy group diving into the massive
  cluster A2142}},
  \href{http://dx.doi.org/10.1051/0004-6361/201424259}{\emph{A\&A} {\bf 570}
  (Oct., 2014) A119}, [\href{http://arxiv.org/abs/1408.1394}{{\tt 1408.1394}}].

\bibitem{2013ApJ...779..189F}
D.~{Farnsworth}, L.~{Rudnick}, S.~{Brown} and G.~{Brunetti}, \emph{{Discovery
  of Megaparsec-scale, Low Surface Brightness Nonthermal Emission in Merging
  Galaxy Clusters Using the Green Bank Telescope}},
  \href{http://dx.doi.org/10.1088/0004-637X/779/2/189}{\emph{ApJ} {\bf 779}
  (Dec., 2013) 189}, [\href{http://arxiv.org/abs/1311.3313}{{\tt 1311.3313}}].

\bibitem{2013A&A...556A...2V}
{van Haarlem}, M.~{Wise} et~al., \emph{{LOFAR: The LOw-Frequency ARray}},
  \href{http://dx.doi.org/10.1051/0004-6361/201220873}{\emph{A\&A} {\bf 556}
  (Aug., 2013) A2}, [\href{http://arxiv.org/abs/1305.3550}{{\tt 1305.3550}}].

\bibitem{2013PASA...30....7T}
S.~J. {Tingay}, R.~{Goeke} et~al., \emph{{The Murchison Widefield Array: The
  Square Kilometre Array Precursor at Low Radio Frequencies}},
  \href{http://dx.doi.org/10.1017/pasa.2012.007}{\emph{Publications of the
  Astronomical Society of Australia} {\bf 30} (Jan., 2013) 7},
  [\href{http://arxiv.org/abs/1206.6945}{{\tt 1206.6945}}].

\bibitem{2009IEEEP..97.1482D}
P.~E. {Dewdney}, P.~J. {Hall}, R.~T. {Schilizzi} and T.~J.~L.~W. {Lazio},
  \emph{{The Square Kilometre Array}},
  \href{http://dx.doi.org/10.1109/JPROC.2009.2021005}{\emph{IEEE Proceedings}
  {\bf 97} (Aug., 2009) 1482--1496}.

\bibitem{2004ApJ...617..281K}
U.~{Keshet}, E.~{Waxman} and A.~{Loeb}, \emph{{Imprint of Intergalactic Shocks
  on the Radio Sky}}, \href{http://dx.doi.org/10.1086/424837}{\emph{ApJ} {\bf
  617} (Dec., 2004) 281--302},
  [\href{http://arxiv.org/abs/astro-ph/0402320}{{\tt astro-ph/0402320}}].

\bibitem{2012SSRv..166..215B}
R.~{Beck}, \emph{{Magnetic Fields in Galaxies}},
  \href{http://dx.doi.org/10.1007/s11214-011-9782-z}{\emph{Space Science
  Reviews} {\bf 166} (May, 2012) 215--230}.

\bibitem{Bleem:2014iim}
{\scshape SPT} collaboration, L.~Bleem et~al., \emph{{Galaxy Clusters
  Discovered via the Sunyaev-Zel'dovich Effect in the 2500-square-degree SPT-SZ
  survey}},
  \href{http://dx.doi.org/10.1088/0067-0049/216/2/27}{\emph{Astrophys.J.Suppl.}
  {\bf 216} (2015) 27}, [\href{http://arxiv.org/abs/1409.0850}{{\tt
  1409.0850}}].

\bibitem{2011ApJ...737...61M}
T.~A. {Marriage}, {Acquaviva} et~al., \emph{{The Atacama Cosmology Telescope:
  Sunyaev-Zel'dovich-Selected Galaxy Clusters at 148 GHz in the 2008 Survey}},
  \href{http://dx.doi.org/10.1088/0004-637X/737/2/61}{\emph{ApJ} {\bf 737}
  (Aug., 2011) 61}, [\href{http://arxiv.org/abs/1010.1065}{{\tt 1010.1065}}].

\bibitem{2013JCAP...07..008H}
M.~{Hasselfield}, M.~{Hilton} et~al., \emph{{The Atacama Cosmology Telescope:
  Sunyaev-Zel'dovich selected galaxy clusters at 148 GHz from three seasons of
  data}}, \href{http://dx.doi.org/10.1088/1475-7516/2013/07/008}{\emph{JCAP}
  {\bf 7} (July, 2013) 8}, [\href{http://arxiv.org/abs/1301.0816}{{\tt
  1301.0816}}].

\bibitem{2014A&A...571A..29P}
{Planck Collaboration}, P.~A.~R. {Ade}, N.~{Aghanim}, C.~{Armitage-Caplan},
  M.~{Arnaud}, M.~{Ashdown} et~al., \emph{{Planck 2013 results. XXIX. The
  Planck catalogue of Sunyaev-Zeldovich sources}},
  \href{http://dx.doi.org/10.1051/0004-6361/201321523}{\emph{Astronomy \&
  Astrophysics} {\bf 571} (Nov., 2014) A29},
  [\href{http://arxiv.org/abs/1303.5089}{{\tt 1303.5089}}].

\bibitem{2011A&A...534A.109P}
R.~{Piffaretti}, M.~{Arnaud}, G.~W. {Pratt}, E.~{Pointecouteau} and J.-B.
  {Melin}, \emph{{The MCXC: a meta-catalogue of x-ray detected clusters of
  galaxies}},
  \href{http://dx.doi.org/10.1051/0004-6361/201015377}{\emph{Astronomy \&
  Astrophysics} {\bf 534} (Oct., 2011) A109},
  [\href{http://arxiv.org/abs/1007.1916}{{\tt 1007.1916}}].

\bibitem{Miniati:2015gga}
F.~Miniati and A.~Beresnyak, \emph{{Self-similar energetics in large clusters
  of galaxies}}, \href{http://dx.doi.org/10.1038/nature14552}{\emph{Nature}
  {\bf 523} (2015) 59}, [\href{http://arxiv.org/abs/1507.01940}{{\tt
  1507.01940}}].

\end{thebibliography}\endgroup

\end{document}